\documentclass{aa}  
\usepackage{comment}
\usepackage{url}
\usepackage[colorlinks,citecolor=blue]{hyperref}\usepackage{graphicx}
\usepackage{natbib}
\usepackage{natbib}
\bibpunct{(}{)}{;}{a}{}{,} % to follow the A&A style
\let\svq"
\catcode`"=\active
\def"{\texttt{\svq}}
\usepackage{txfonts}
\usepackage{float}
\begin{document}

   \title{Rotation curves in protoplanetary disks with thermal stratification}

   \subtitle{Physical model and observational evidence in MAPS disks}

   \author{P. Martire\inst{1,2}, C. Longarini\inst{1,8}, G. Lodato\inst{1}, G. P. Rosotti\inst{1}, A. Winter\inst{3}, S. Facchini\inst{1}, C. Hardiman\inst{4}, M. Benisty\inst{3,5}, J. Stadler\inst{3,5},  A. F. Izquierdo\inst{2} and Leonardo Testi\inst{6,7}.
          }

   \institute{Dipartimento di Fisica, Università degli Studi di Milano, Via Celoria 16, Milano, I-20133, Italy\\
              \email{martire@strw.leidenuniv.nl}\\
              \email{cl2000@cam.ac.uk}
    \and{Leiden Observatory, Leiden University, P.O. Box 9513, 2300 RA Leiden, The Netherlands}
    \and Laboratoire Lagrange, Université Côte d’Azur, CNRS, Observatoire de la Côte d’Azur, 06304 Nice, France
    \and School of Physics and Astronomy, Monash University, Clayton, VIC 3800, Australia
    \and Univ. Grenoble Alpes, CNRS, IPAG, 38000 Grenoble, France
    \and{Alma Mater Studiorum Università di Bologna, Dipartimento di Fisica e Astronomia (DIFA), Via Gobetti 93/2, 40129 Bologna, Italy}
    \and{INAF-Osservatorio Astrofisico di Arcetri, L.go E. Fermi 5, I-50125 Firenze, Italy}
    \and{Institute of Astronomy, University of Cambridge, Madingley Road, Cambridge, CB3 0HA, United Kingdom}
             }

   \date{Received ; accepted }
\titlerunning{ Rotation curves in stratified disks}
\authorrunning{Martire et al.}
% \abstract{}{}{}{}{} 
% 5 {} token are mandatory

  \abstract
   {In recent years, the gas kinematics probed by molecular lines detected with ALMA  has opened a new window into the of study protoplanetary disks. High spatial and spectral resolution observations have revealed the complexity of protoplanetary disk structure. Drawing accurate interpretations of these data allows us to better comprehend planet formation} 
   {We investigate the impact of thermal stratification on the azimuthal velocity of protoplanetary disks. High-resolution gas observations reveal velocity differences between CO isotopologues, which cannot be adequately explained with vertically isothermal models. The aim of this work is to determine whether a stratified model can explain this discrepancy.}
   {We analytically solved the hydrostatic equilibrium for a stratified disk and we derived the azimuthal velocity. We tested the model with SPH numerical simulations and then we used it to fit for the star mass, disk mass, and scale radius of the sources in the MAPS sample. In particular, we used $^{12}$CO and $^{13}$CO datacubes. }  
  % We made analytical calculations to model protoplanetary disks considering their vertical gradient of temperature.
%   After having tested our equations in SPH simulations with the code \textsc{phantom}, we implemented our model in MCMC fitting procedures in order to fit physical parameters of the disks around GM Aur and HD 163296.}
   {%By considering the thermal stratification, we are able to reconcile the inconsistency between CO isotopologues rotation curves and to obtain a solid estimate of disk parameters, with a more realistic estimate of the best-fit parameters. 
   When thermal stratification is taken into account, it is possible to reconcile most of the inconsistencies between rotation curves of different isotopologues. A more accurate description of the CO rotation curves offers a deeper understanding of the disk structure. The best-fit values of star mass, disk mass, and scale radius become more realistic and more in line with previous studies. In particular, the quality of the scale radius estimate significantly increases when adopting a stratified model. In light of our results, we computed the gas-to-dust ratio and the Toomre Q parameter. Within our hypothesis, for all the sources, the gas-to-dust ratio appears higher but still close to the standard value of 100 (within a factor of 2). The Toomre Q parameter suggests that the disks are gravitationally stable $(Q>1)$. However, the systems that show spirals presence are closer to the conditions of gravitational instability ($Q\sim 5$).}
   % The value that we obtained for the disk mass and the taper radius are consistent with thermo-chemical models, highlighting that thermal stratification needs to be taken into account to properly interpret rotation curves to determine the self-gravity term}
  {}

   \keywords{protoplanetary disks -- hydrodynamics
                -- accretion, accretion disks
               }

   \maketitle
%
%-------------------------------------------------------------------

\section{Introduction}
Our understanding of the physical properties of protoplanetary disks has improved in recent years thanks to the Atacama Large Millimeter Array (ALMA) \citep{HLTAU}. High spectral and spatial resolution gas observations enable us to probe density, temperature, and velocity fields of protostellar disks, gaining unique information about their structure \citep{MAPSIV,calahan21,teague22,miotello2022,pinte2022, LodatoLongarini}. More recently, the large program Molecules with ALMA at Planet-forming Scales (MAPS) \citep{MAPSI} targeted five protoplanetary disks (MWC 480, IM Lup, GM Aur, HD 163296, and AS 209) in several molecular lines. For optically thick line emission, the gas temperature can be measured along the emission surface directly from the peak surface brightness of the channel maps \citep{MAPSIV}. Given the varying heights of these emitting layers surfaces, it is possible to infer the thermal structure in disks, proving the existence of a vertical thermal stratification in them \citep{Dartois, Rosenfeld2013, Pinte18}, as expected from basic radiative transfer arguments \citep{chiang97,dalessio98,dalessio99}. Although disk models have usually been considered as vertically isothermal, the vertical gradient of temperature leads to considerable corrections in the calculation of density structure and azimuthal velocity, which results in several percent deviations from the Keplerian velocity \citep{Rosenfeld2013}. Accounting for such differences is important not only to infer stellar masses, but also to accurately constrain the disk pressure structure and disk mass. Such parameters are of great importance with respect to interpreting velocity deviations, which may serve as signposts for planets \citep{Pinte18,rabago21,bollati21,discminer1,bae21}, dust trapping \citep{teague18a, Rosotti2020} or disk instabilities \citep{hall20,Terry21,Longarini21,barraza21}.

In this paper, we analytically derive the density and velocity field of protostellar disks with thermal stratification, generalizing the work of \citet{TakeuchiLin2002}. A similar analysis of MAPS data with vertically isothermal disks was performed by \cite{LodatoLongarini}. We test the model against hydrodynamical simulations and we apply it to the whole MAPS sample for $^{12}$CO and $^{13}$CO data. In Sect. \ref{S1}, we present the model, solving the vertical hydrostatic equilibrium and obtaining an expression for the azimuthal velocity. In Sect. \ref{S2}, we present the numerical setup and a comparison between the model and simulations. In Sect. \ref{S3}, we apply the model and we discuss our findings. Finally, in Sect. \ref{S5}, we compute the gas-to-dust ratio and the Toomre Q parameter, and we draw our conclusions. 

\section{Model}\label{S1}
\subsection{Assumptions}
In our analytical calculations, we did not make any assumption on the surface density $\Sigma$, considering it as arbitrary. However,  to apply the model to observations, we were forced to choose a parameterization for the surface density and we assume that it is described by the self-similar solution from \citet{Lind&Prin}:
\begin{equation}
\label{eq: surf density}
    \Sigma = \frac{(2-\gamma)M_\text{d}}{2\pi R_\text{c}^2} \Bigg(\frac{R}{R_ \text{c}}\Bigg)^{-\gamma}\exp\Bigg[-\Bigg(\frac{R}{R_\text{c}}\Bigg)^{2-\gamma}\Bigg],
\end{equation}
where $M_\text{d}$  and $ R_\text{c}$ are the disk mass and the scale radius respectively; $R$ is the cylindrical radius and $\gamma$ is a free parameter describing the steepness of the surface density. The disk density at the midplane $\rho_\text{mid}$ is:
\begin{equation}
\label{density mid}
\rho_\text{mid} = \frac{\Sigma}{\sqrt{2\pi}H_\text{mid}}\propto R^{-(\gamma+(3-q)/2)}\exp\Bigg[-\Bigg(\frac{R}{R_\text{c}}\Bigg)^{2-\gamma}\Bigg],
\end{equation}
where $H_\text{mid}=c_{\text{s},\text{mid}}/\Omega_\text{k}$ is the typical scale height of the disk at the midplane, $c_{\text{s},_\text{mid}} =\sqrt{k_\text{b}T_\text{mid}/(\mu m_\text{p})}\propto R^{-q/2}$ is the sound speed at the disk midplane, $k_\text{b}$ is the Boltzmann constant, $T_{\text{mid}} = T_{\text{mid},100}({R}/{100\text{au}})^{-q}$  is the temperature at midplane, $\mu$ is the mean molecular weight (usually assumed to be 2.1), $m_\text{p}$ is the proton mass, and $\Omega_\text{k}=\sqrt{GM_\star/R^3}$ is the Keplerian angular velocity ($G$ is the gravitational constant and $M_\star$ is the stellar mass).

From the literature \citep{chiang97, Dullemond} and observational data \citep{Rosenfeld2013, Pinte18, MAPSIV}, we know that protoplanetary disks are thermally stratified. We take this into account by defining a function, $f$, that describes the dependency of the temperature, $T$, on height, such that:
\begin{gather}
    \label{temp} T(R,z)=T_\text{mid}(R)f(R,z),\\ c_\text{s}^2(R,z)=c_{\text{s},\text{mid}}^2(R)f(R,z).
\end{gather} 
We underline that the isothermal case can be obtained considering $f\equiv1$, thus $T=T_\text{mid}(R)$.
%In this context, it is useful to define the atmosphere temperature, i.e. the temperarure at the disk surface. Both the midplane and the atmosphere temperatures are described as power law with the radius
%\begin{gather}
 %   T_\text{mid}(R) = T_\text{mid,100}\Bigg(\frac{R}{100\text{au}}\Bigg)^{-q_\text{mid}},\\
  %  T_{\text{atm}}(R) = T_{\text{atm},100}\Bigg(\frac{R}{100\text{au}}\Bigg)^{-q_\text{atm}},
%\end{gather}
As for the density, we assume that: 
\begin{equation}
\label{rho}
    \rho= \rho(R,z)=\rho_\text{mid}(R)g(R,z),
\end{equation}
where $g$  describes how the density changes vertically. 
We note that in order to smoothly connect the functions above to their value at midplane it is necessary that $f(z=0)=1=g(z=0)$. Assuming a barotropic fluid, the pressure, $P$, is given by:
\begin{equation}
    P(R,z)=P_\text{mid}(R)fg(R,z)=c_{\text{s},\text{mid}}^2(R)\rho_\text{mid}(R)fg(R,z).
\end{equation}
While the profile of $f$ is arbitrary, this does not hold for $g$, whose value is set by solving the vertical hydrostatic equilibrium.

\subsection{Hydrostatic equilibrium and rotation curve}\label{subsec: modelling}
To compute the vertical density profile we assume a non-self-gravitating disk under the condition of hydrostatic equilibrium in the vertical direction:
\begin{equation}\label{hydro_eq}
    \frac{1}{\rho}\frac{dP}{dz}=-\frac{d\Phi_\star}{dz},
\end{equation}
where $\Phi_\star=-GM_\star/r$ is the stellar potential ($r=\sqrt{R^2+z^2}$ is the spherical radius). Equation \eqref{hydro_eq} can be written as (for further details, see Appendix \ref{calc}): 
\begin{equation}
    c_{\text{s},_\text{mid}}^2f\frac{d\log(fg)}{dz} = -\Omega_\text{k}^2 z\Bigg[1+\Bigg(\frac{z}{R}\Bigg)^2\Bigg]^{-3/2}.  
\end{equation}
Solving for $\log(fg)$, we find: 
\begin{equation}  
\label{fg}
    \log(fg) = -\frac{1}{H_\text{mid}^2}\int_0^z \frac{z'}{f}\left[1+\left(\frac{z'}{R}\right)^2\right]^{-3/2} dz',
\end{equation}
and hence the density is given by: 
\begin{equation}\label{density}
    \rho(R,z) = \frac{\rho_\text{mid}(R)}{f(R,z)}\exp\left\{-\frac{1}{H_\text{mid}^2}\int_0^z \frac{z'}{f(R,z')}\left[1+\left(\frac{z'}{R}\right)^2\right]^{-3/2} dz'\right\}.
\end{equation}
Assuming the condition of centrifugal balance, the rotation curve is given by the radial component of Navier-Stokes equation:
\begin{equation}
    \label{eq:NS}
    v_\phi^2 (R,z) =\frac{R}{\rho}\frac{dP}{dR}(R,z) + R\frac{d\Phi_\star}{dR}(R,z).
\end{equation}
The first term in Eq.\eqref{eq:NS} can be written as (for further details, see Appendix \ref{calc})
\begin{equation}
\label{eq: grad P}
         \frac{R}{\rho}\frac{dP}{dR} = c_{\text{s},_\text{mid}}^2f\Bigg[\frac{d\log P_\text{mid}}{d\log R}+\frac{d\log(fg)}{d\log R}\Bigg],
\end{equation}
and the second one as
\begin{equation}
\label{eq:pot}
    R\frac{d\Phi_\star}{dR} (R,z) = v_\text{k}^2\left[1+\left(\frac{z}{R}\right)^2\right]^{-3/2},
\end{equation}
where $v_\text{k}=\sqrt{GM_\star/R}$ is the Keplerian velocity. Therefore, the rotation curve is
\begin{equation}\label{rotationcurve_strat_general}\begin{split}
    v_\phi^2(R,z) = v_\text{k}^2 \left\{\left[1+\left(\frac{z}{R}\right)^2\right]^{-3/2} + \left[\frac{d\log P_\text{mid}}{d\log R}  + \right. \right. \\
    \left.\left. + \frac{\text{d}\log(fg)}{\text{d}\log R}\right]\left(\frac{H}{R}\right)_\text{mid}^2 f(R,z)   \right\}.
\end{split}\end{equation}
In the self-similar case, this becomes:
\begin{equation}\label{rotationcurve_strat}\begin{split}
    v_\phi^2(R,z) = v_\text{k}^2 \left\{\left[1+\left(\frac{z}{R}\right)^2\right]^{-3/2} - \left[\gamma^\prime + (2-\gamma)\left(\frac{R}{R_c}\right)^{2-\gamma}  - \right. \right. \\ \left.\left. - \frac{\text{d}\log(fg)}{\text{d}\log R}\right]\left(\frac{H}{R}\right)_\text{mid}^2 f(R,z)   \right\},
\end{split}\end{equation}
where $\gamma^\prime = \gamma + (3+q)/2 $. Each term of Eq. \eqref{rotationcurve_strat} can be easily interpreted: $[1+(z/R)^2]^{-3/2}$ is the star contribution at the height $z$, $\gamma^\prime$ is the effect of the power law scaling of the pressure, $(2-\gamma)(R/R_c)^{2-\gamma}$ is the effect of the exponential truncation and the logarithmic term is the effect of the vertical stratification. Since the latter is the derivative of a product, we do not know its sign a priori; thus, we also do not know if the rotation is accelerated or slowed down by thermal stratification (see Appendix \ref{calc}). In any case, in all our attempts this term never dominates over the variation of gravity with $z$. Thus,  we found rotation to slow down with z and this effect is more pronounced as compared to the isothermal case when considering the parameters of the MAPS sample.

\begin{figure*}[h!tbp]
    \centering
    \includegraphics[scale = 0.45]{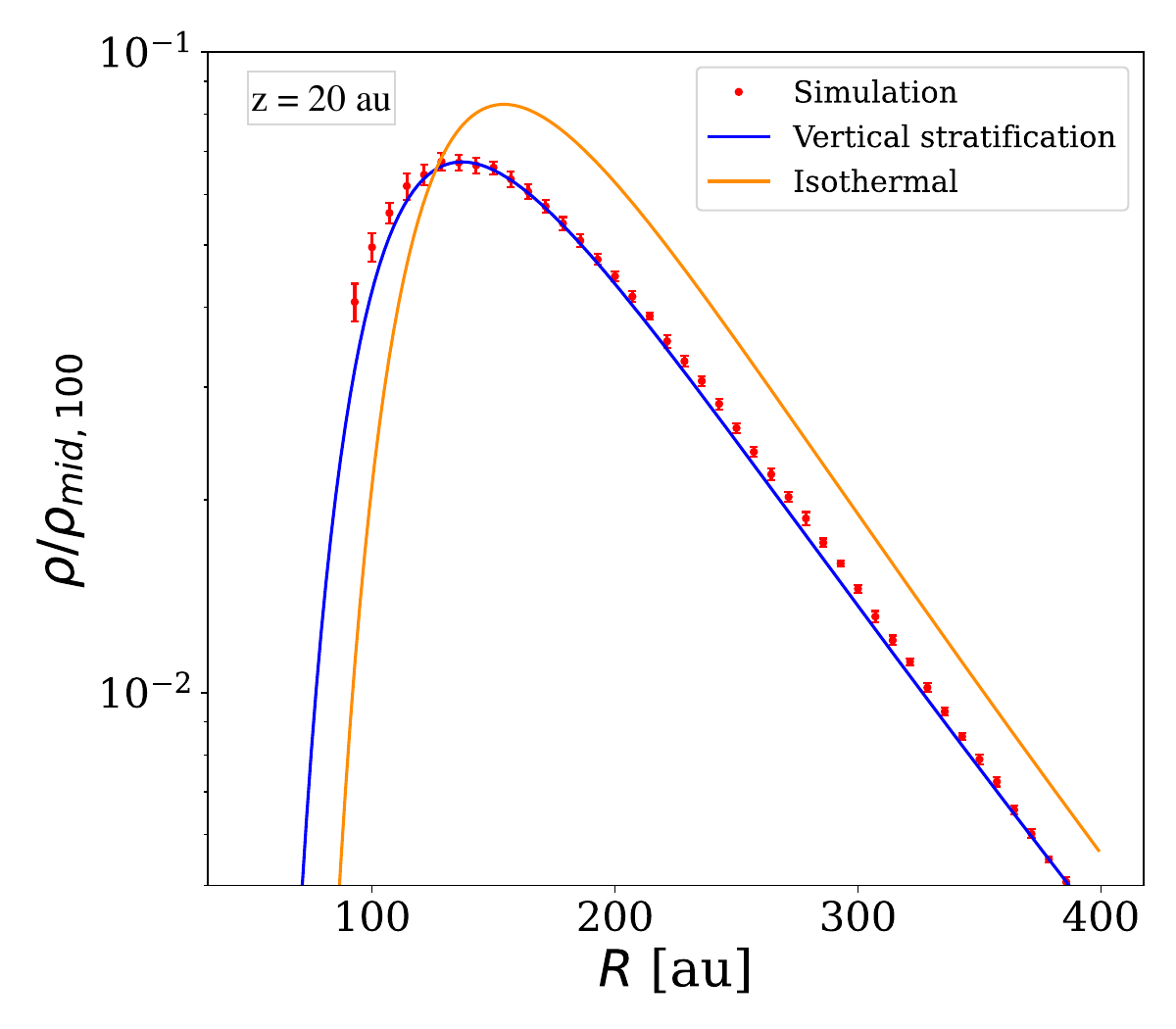}
    \includegraphics[scale = 0.45]{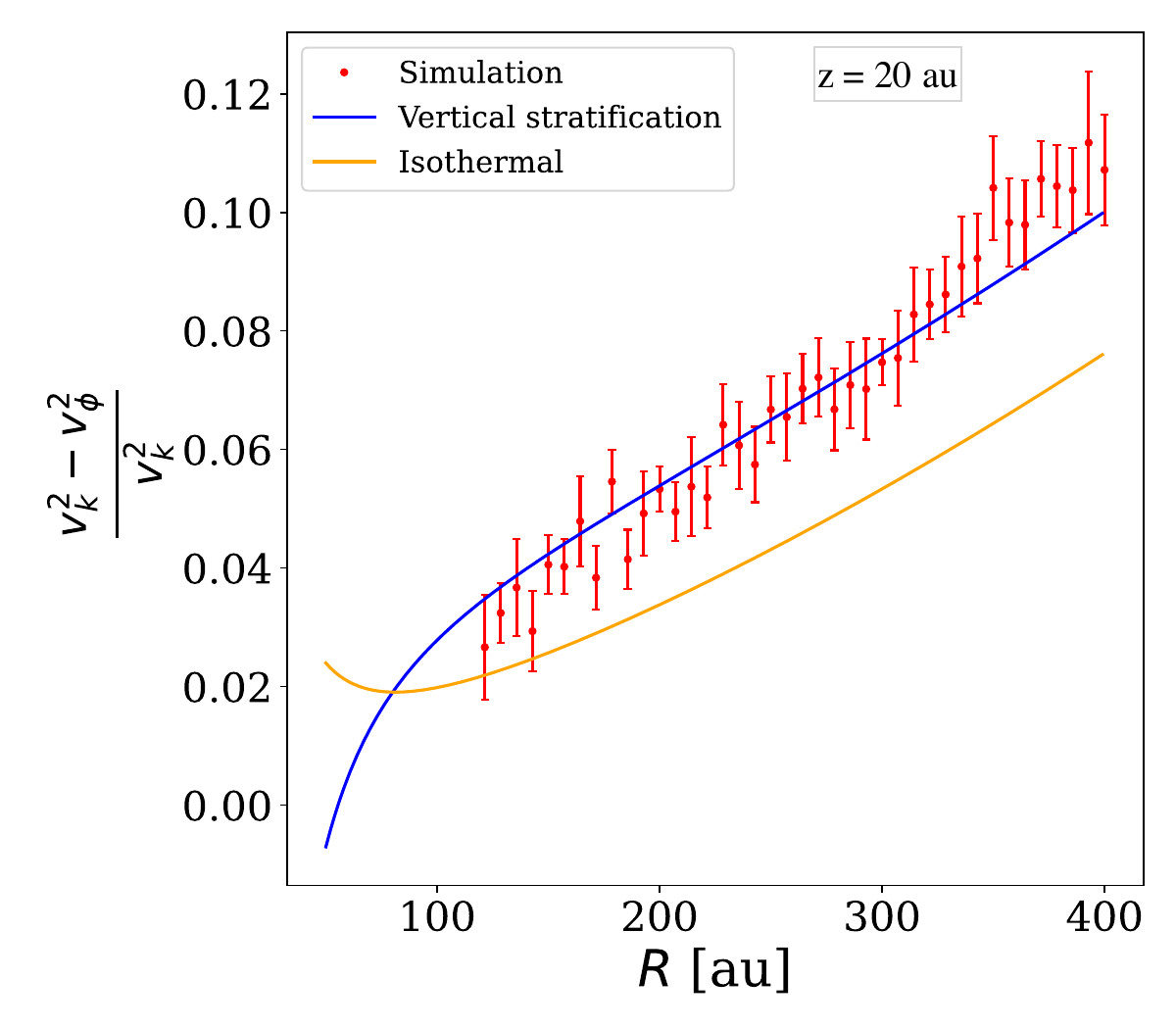}
    \caption{Comparison between the analytical models and the simulation after eight outer orbits and at $z=20$au. Left panel: Density field. Right panel: Pressure gradient term of the rotation curve.  The red dots represent the simulation data, while the blue and orange line show the thermally stratified model and the isothermal one, respectively. The model with thermal stratification matches very well the simulation.}
    \label{simulation}
\end{figure*} 

We underline that for the isothermal case ($f\equiv1$), this expression reduces to the one derived and analyzed by \citet{LodatoLongarini}, while Eq. \eqref{fg} is simplified as:
\begin{equation}
    \log g = -\frac{1}{H_\text{mid}^2}\int_0^z \frac{z'dz'}{[1+(z'/R)^2]^{3/2}} = -\frac{R^2}{H_\text{mid}^2} \left(1-\frac{1}{\sqrt{1+z^2/R^2}}   \right).
\end{equation}
Therefore, the density in the isothermal case is given by:
\begin{equation}
    \rho(R,z) = \rho_\text{mid}(R)\exp\left[\frac{R^2}{H_\text{mid}^2}\left(\frac{1}{\sqrt{1+z^2/R^2}}-1\right)\right].
\end{equation}
If the disk is self-gravitating, we should add to the right-hand side of Eq.\eqref{eq:NS} the self-gravitating term \citep{BertinLodatoselfgrav}:
\begin{equation}
\begin{gathered}
	\frac{R}{\rho}\frac{d\Phi_\text{d}}{dR}(R,z) = G \int^\infty_0 \Bigg[K(k) - \frac{1}{4}\Bigg(\frac{k^2}{1-k^2}\Bigg)\times \\
    \Bigg(\frac{r}{R}-\frac{R}{r}+\frac{z^2}{Rr}\Bigg) E(k)\Bigg]\sqrt{\frac{r}{R}} k\Sigma(r) dr,
\end{gathered}
\end{equation}
where $K(k)$ and $E(k)$ are complete elliptic integrals \citep{abramowitz} and $k^2 =  4Rr/[(R + r)^2 + z^2]$. 

\subsection{Temperature prescriptions}
\label{temp prescrp}
The two parameterizations of the vertical temperature more often used are given by \citet{Dartois} and \citet{Dullemond}. In this work we will use the one by \citet{Dullemond}, which is given by:
\begin{equation}\label{Dull}
    T(R,z)^4 = T_{\epsilon}^4(R) +\frac{1}{2}T_\text{atm}^4(R)\Bigg[1+\tanh\Big(\frac{z}{Z_q(R)}-\alpha\Big)\Bigg]
\end{equation}
and, thus, 
\begin{equation}\label{Dull f}
        f(R,z) = \Bigg\{\left(\frac{T_\epsilon}{T_\text{mid}}\right)^4+\frac{1}{2}\left(\frac{T_\text{atm}}{T_\text{mid}}\right)^4(R)\Bigg[1+\tanh\Big(\frac{z}{Z_q(R)}-\alpha\Big)\Bigg]\Bigg\}^{1/4},
\end{equation}
where the atmospheric temperature is parameterized  as $T_{\text{atm}}(R) = T_{\text{atm},100}({R}/{100\text{au}})^{-q_\text{atm}}$, $T_\epsilon$ is considered as an approximation of the temperature at the midplane $T_\epsilon \simeq T_\text{mid}$. $Z_q(R)$ is defined as $Z_q(R)=\zeta_{100}(R/100\text{au})^\beta$ and $\alpha$ is a parameter that describes where the transition from midplane to atmospheric temperature occurs in the vertical direction. We note that in this case, $f(R,z = 0)\neq1$ and, thus, the temperature does not smoothly connect to its value at midplane. We discuss this in Appendix \ref{Dullcorrection}, but we underline that Eq. \eqref{Dull} is a good approximation for the five disks within the MAPS large program in most of the radial extent of the disk.

%In this work, we approximate $T_\epsilon \simeq T_\text{mid}$, even though the two quantities may be different. We discuss this in appendix \ref{Dullcorrection}, but we underline that Eq. \eqref{Dull} is a good approximation for the five disks within the MAPS large program in most of the radial extent of the disk.

Once the function $f$ is defined, Eqs. \eqref{density} and \eqref{rotationcurve_strat_general} can be solved semi-analytically and entirely specify the rotation curve. We have implemented this calculation in \textsc{DYSC}\footnote{The code is publicly available at \url{https://github.com/crislong/DySc}}.

\section{Comparison with numerical simulations}\label{S2}
In this work, we performed numerical smoothed particle hydrodynamics (SPH) simulations of protostellar disks using the \textsc{phantom} code \citep{phantom18}. This code is widely used in the astrophysical community to study gas and dust dynamics in accretion disks \citep{Dipierro15,Ragusa17,Curone22} and it has recently been employed in kinematical studies \citep{Pinte18,hall20,Terry21,Verrios22}. The aim of this simulation is to test the model before applying it to actual data. 

To test the analytical model, we simulated a thermally stratified disk using the parameters of MWC 480 presented in \cite{MAPSIV}. The simulation has been performed with $N=10^6$ gas particles, initially distributed as a tapered power law density profile, smoothed at the inner radius, with $\gamma =1$ and $R_c= 150$au, between $R_\text{in}=10$au and $R_\text{out}=400$au. The mass of the star is $2.1 \text{M}_\odot$. For the temperature structure we used the Dullemond prescription given by Eq. \eqref{Dull}, with $\zeta_0=7 \text{au}, \alpha = 2.78, \beta=-0.05, T_{\text{mid},100}=27$K, $q=0.23$, $T_{\text{atm,100}}=69$K, and $q_{\text{atm}}=0.7$. The $\alpha_{\text{SS}}$ Shakura \& Sunyaev \citep{shakurasunyaev} viscosity coefficient has been set to 0.005. No self-gravity or dust have been included in the simulation.
 
We let the system evolve and reach hydrostatic equilibrium. We observed that after a couple of orbits the system reaches a relaxed state. We decided to analyze the output of the simulation after eight outer orbits $(\sim 45\text{kyr})$. In Fig. \ref{simulation}, we show a comparison between the density and the velocity of the simulations (red dots) at $z=20$ au and both the isothermal and stratified model predictions. The red dots represent the azimuthal average of the respective quantity computed by averaging over all SPH particles within each of the 50 radial bins and the error bar is the corresponding standard deviation. 
Since we are plotting quantities at z = 20 au, we have excluded the inner points because at those radii (R < 100 au) the disk has a smaller hydrostatic height H, causing numerical issues in the resolution of our simulation.
The stratified model perfectly describes the density and the velocity field of the simulation and is a significant improvement over the isothermal one. In particular, in the right panel of Fig. \ref{simulation}, we can see that the difference between the azimuthal velocity and the Keplerian velocity $(v_\text{k}^2-v_\phi^2)/v_\text{k}^2$ reaches the $10-12\%$ and only the stratified model is able to reproduce it.

\section{Applying the model}\label{S3}
\begin{figure*}
    \centering
    \includegraphics[width=\columnwidth]{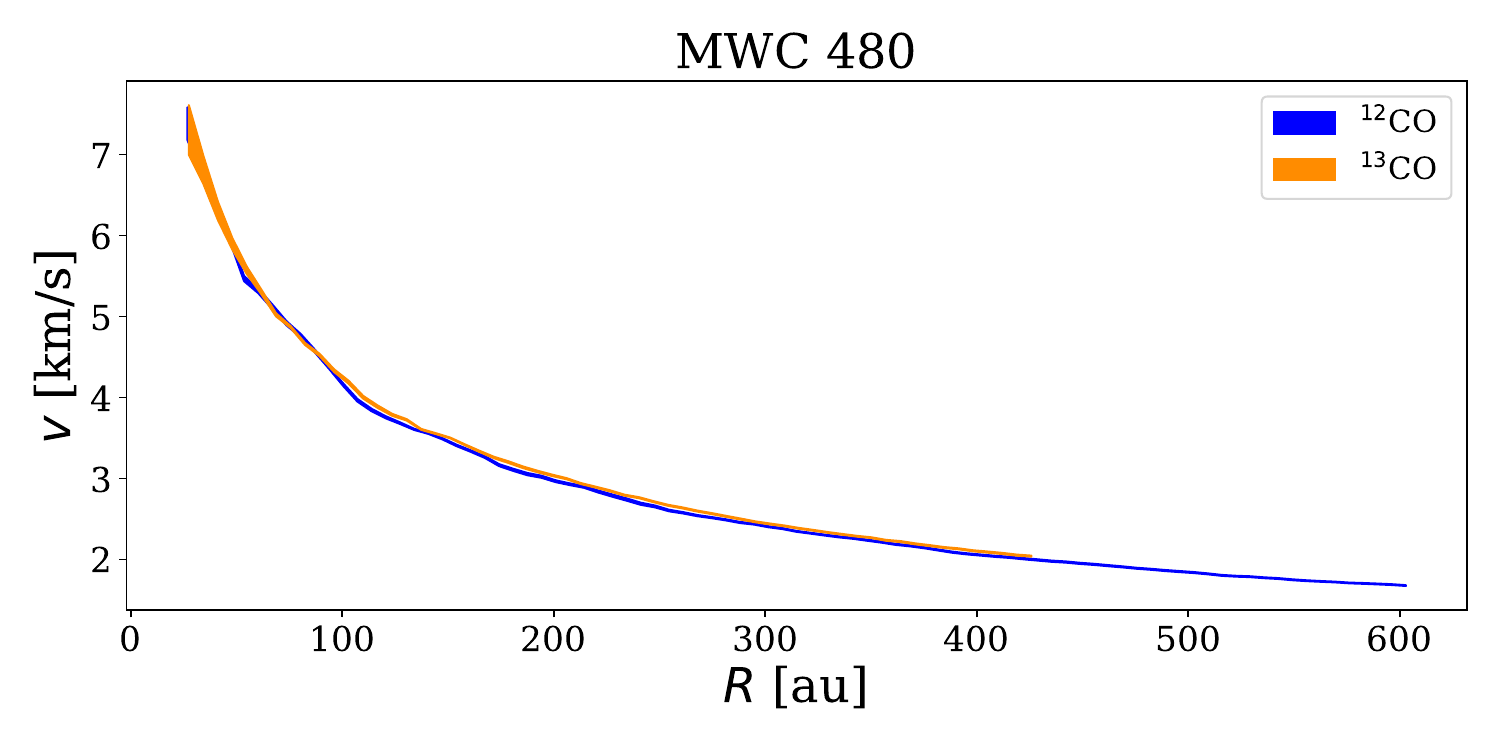}
    \includegraphics[width=\columnwidth]{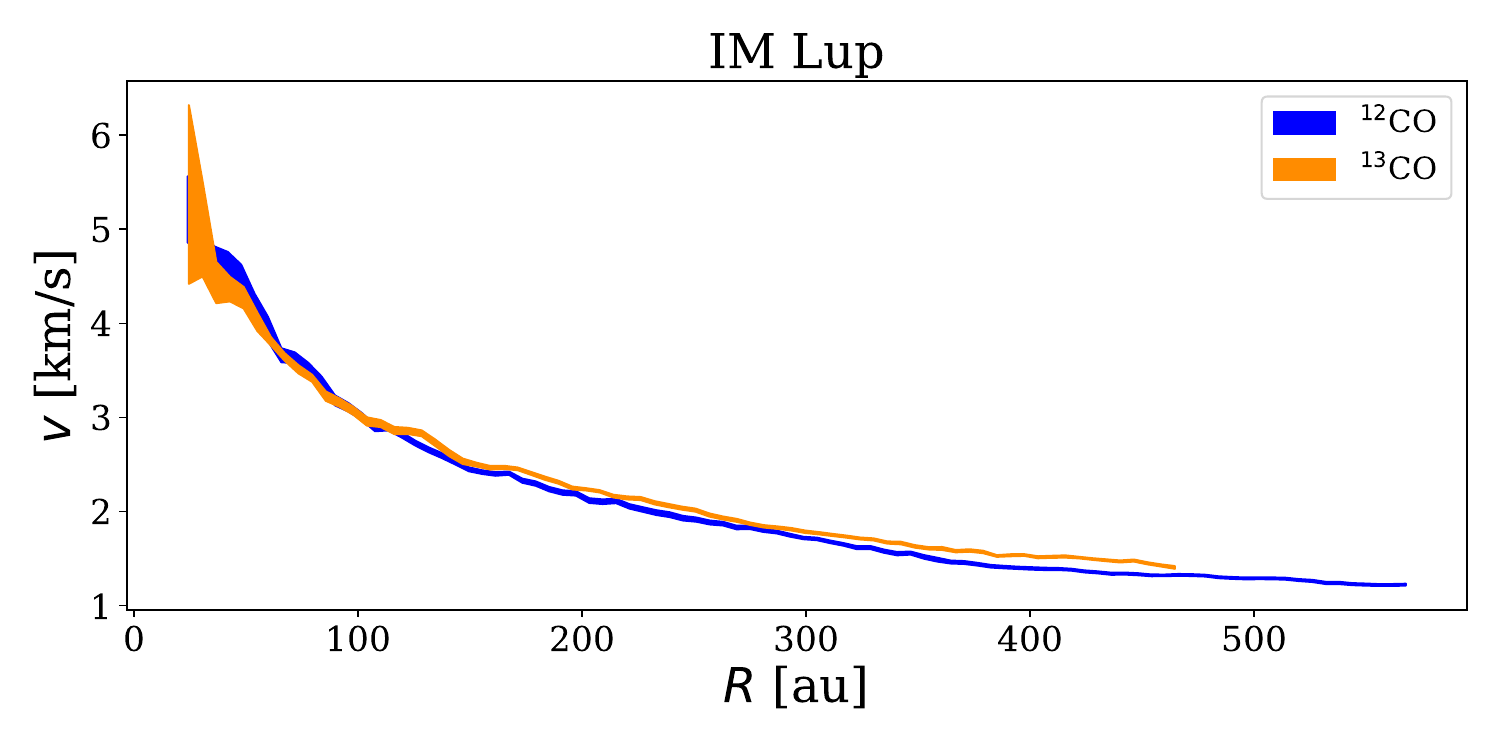}
    \includegraphics[width=\columnwidth]{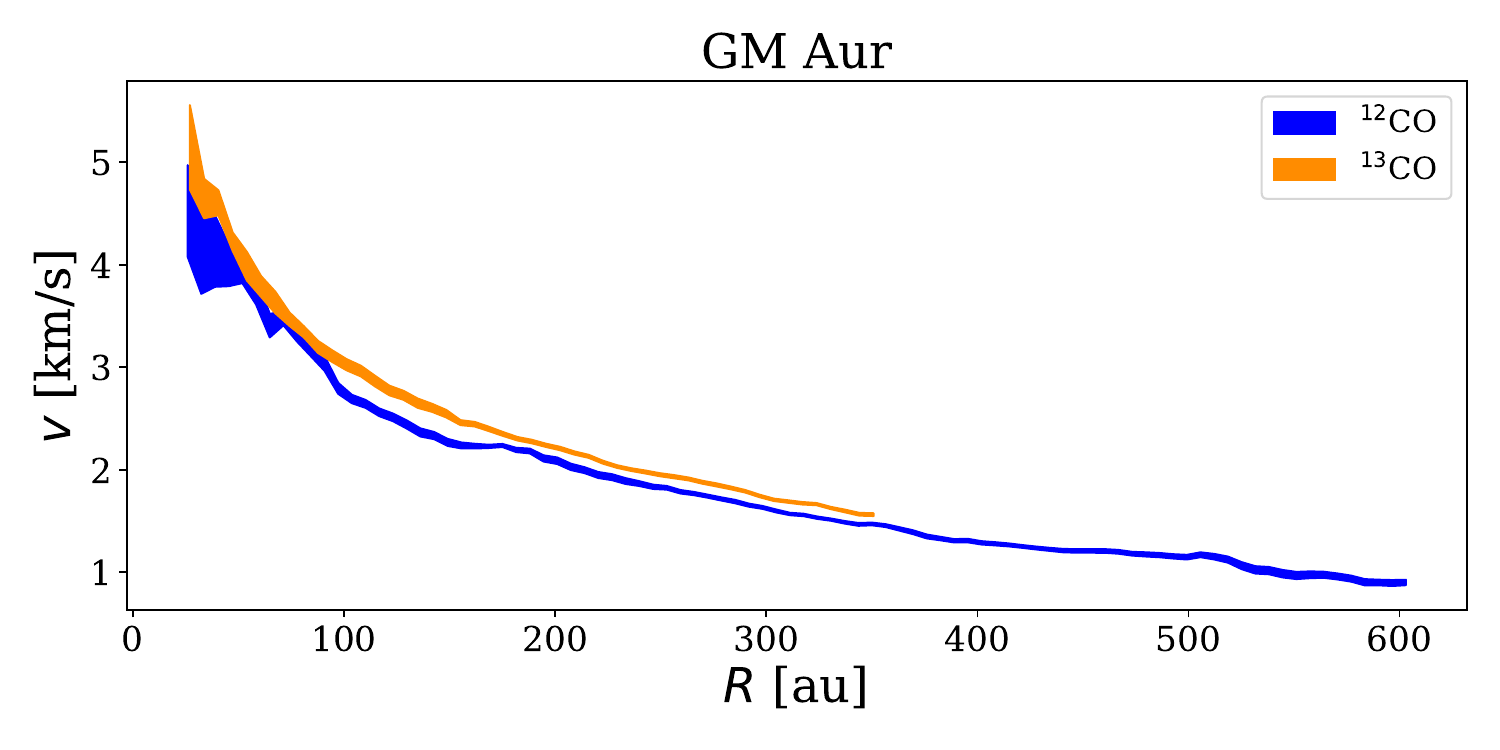}
    \includegraphics[width=\columnwidth]{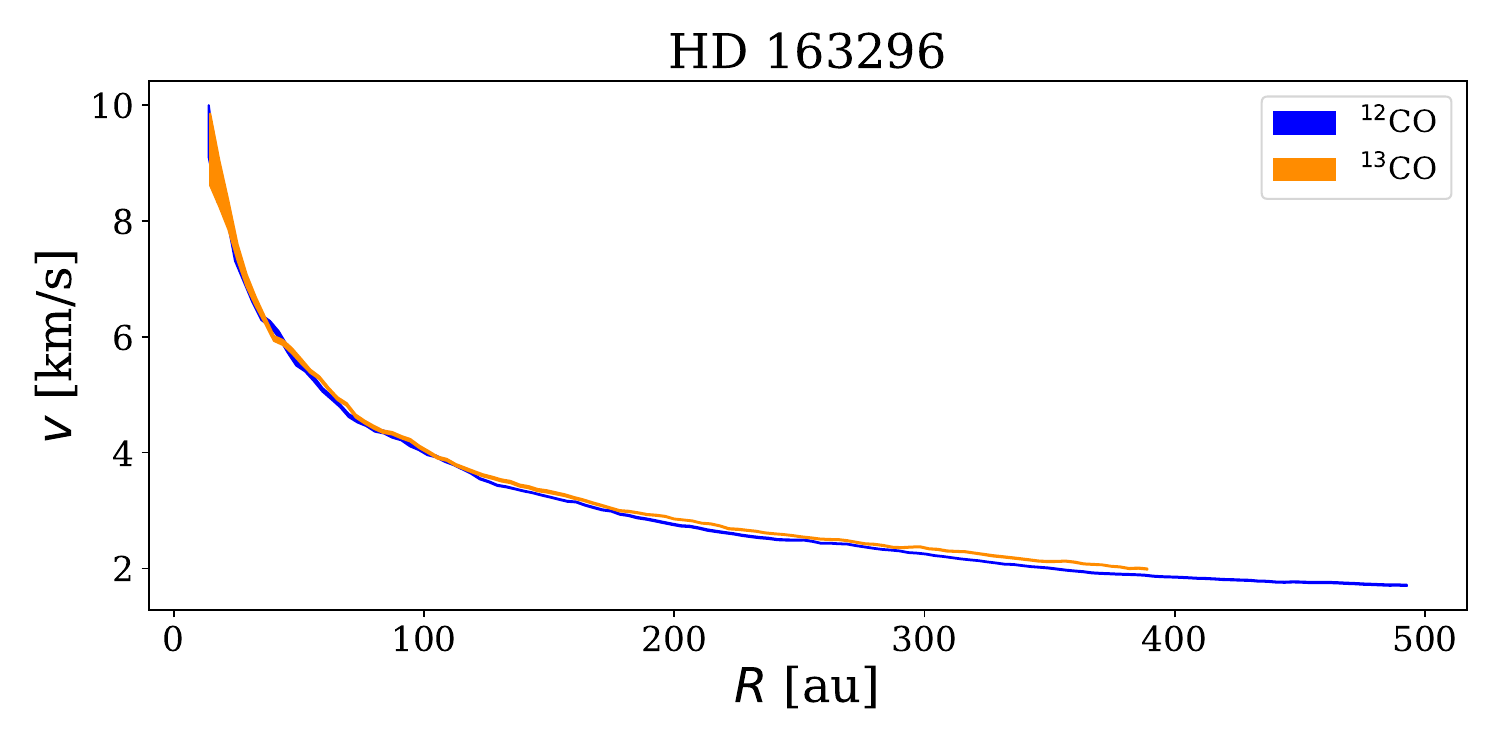}
    \includegraphics[width=\columnwidth]{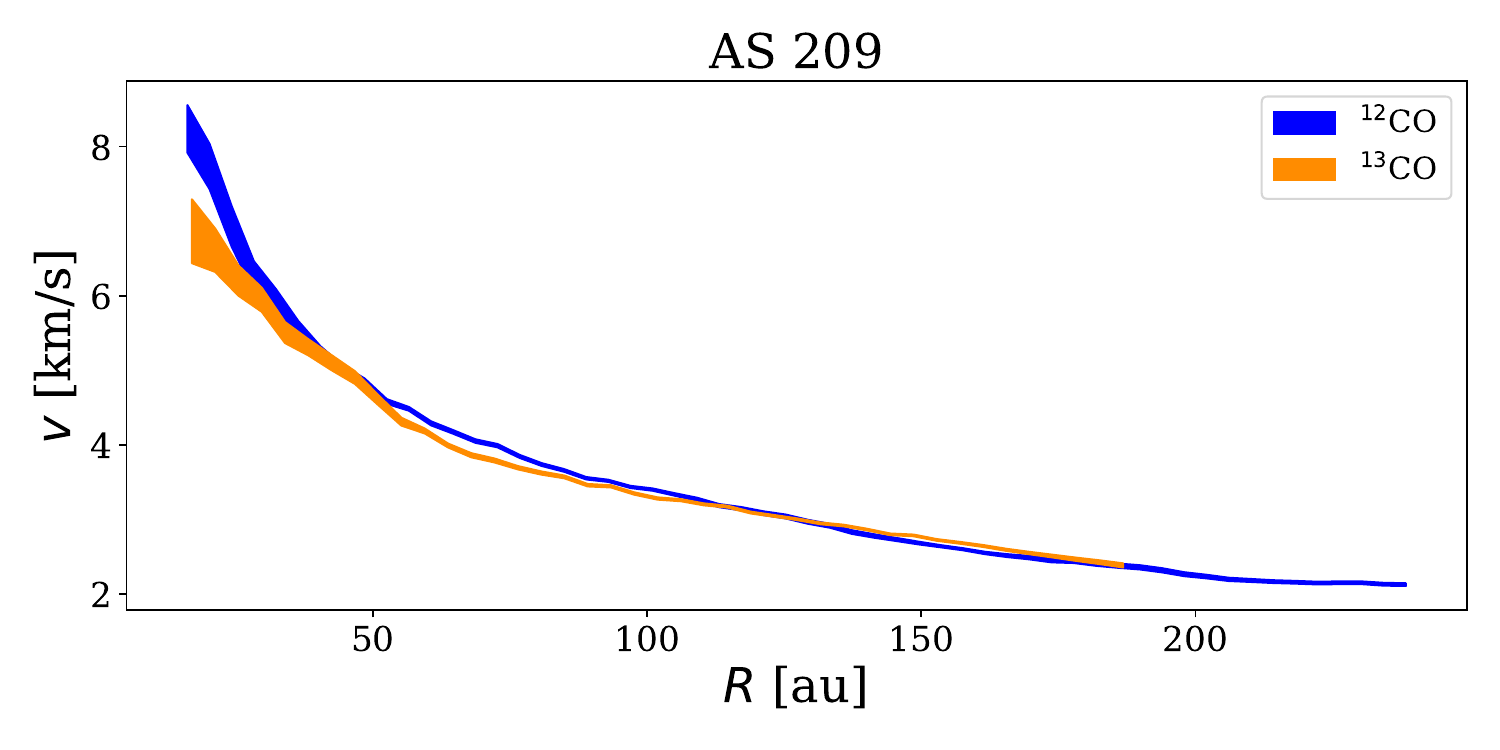}

    \caption{$^{12}$CO and $^{13}$CO rotation curves of the MAPS disks extracted with \textsc{discminer}.}
    \label{rcurves_1213}
\end{figure*}

\subsection{The curves}
\begin{figure*}[h!tbp]
    \centering
    \includegraphics[scale=0.435]{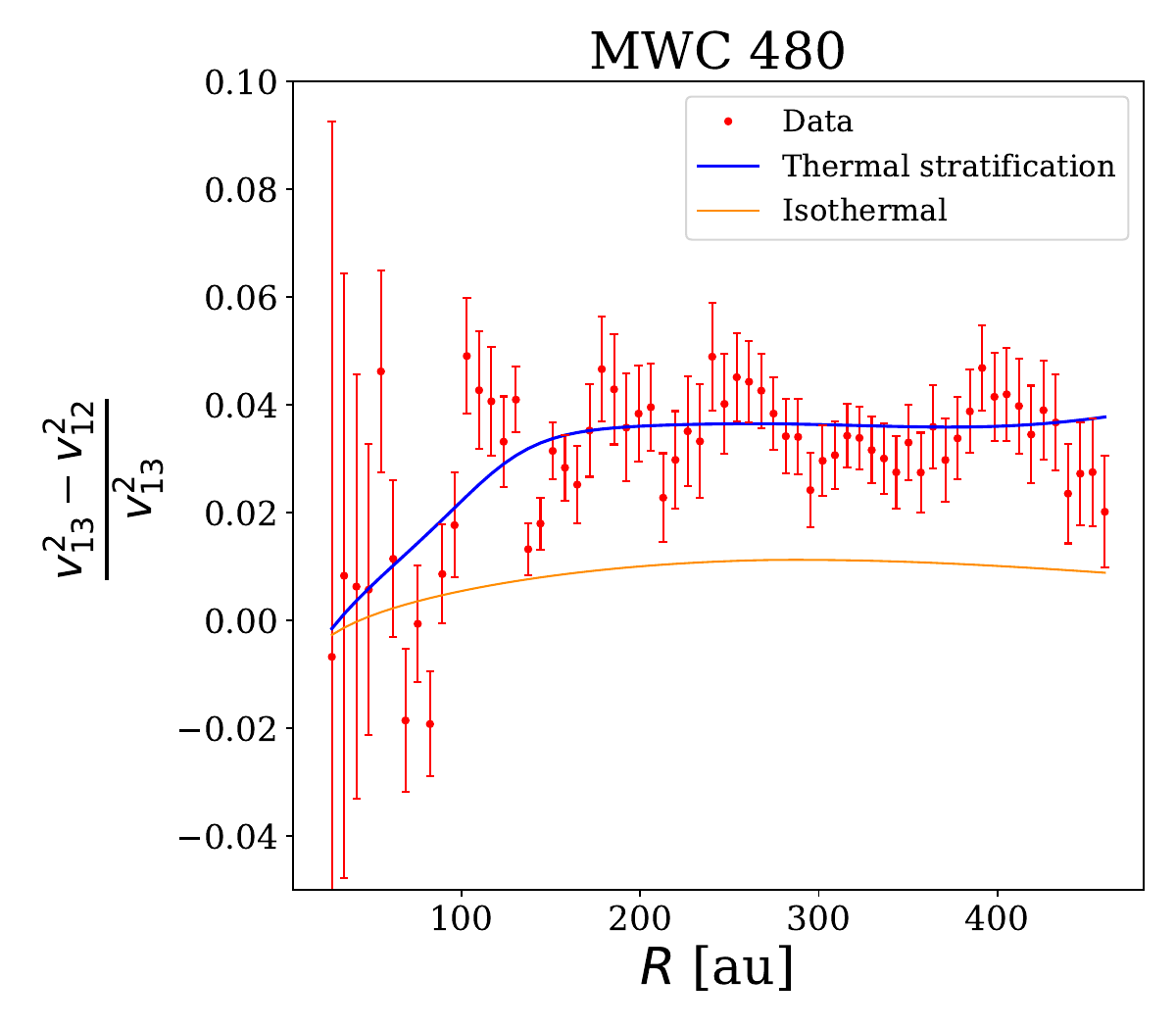}
    \includegraphics[scale=0.435]{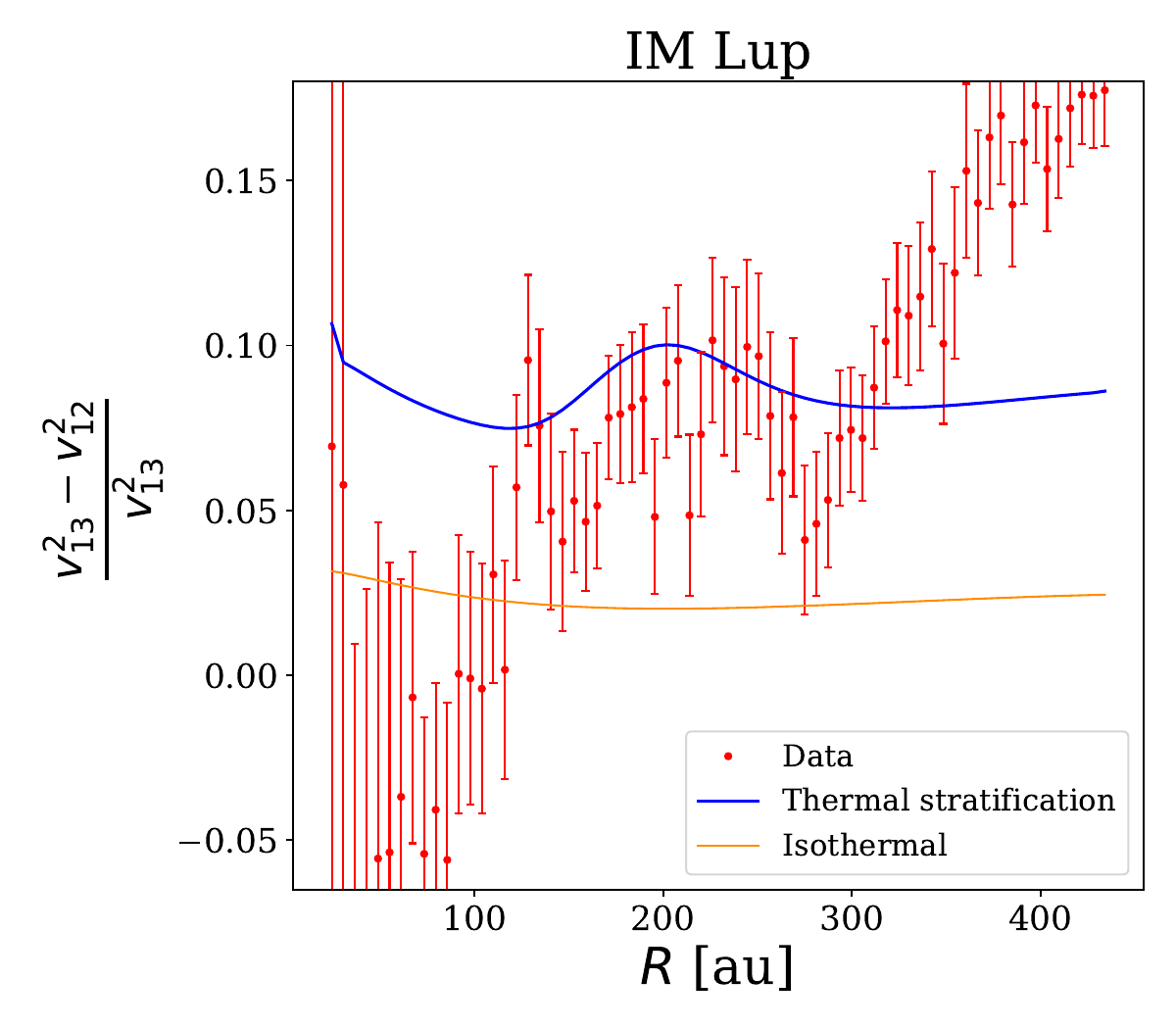}
        \includegraphics[scale=0.435]{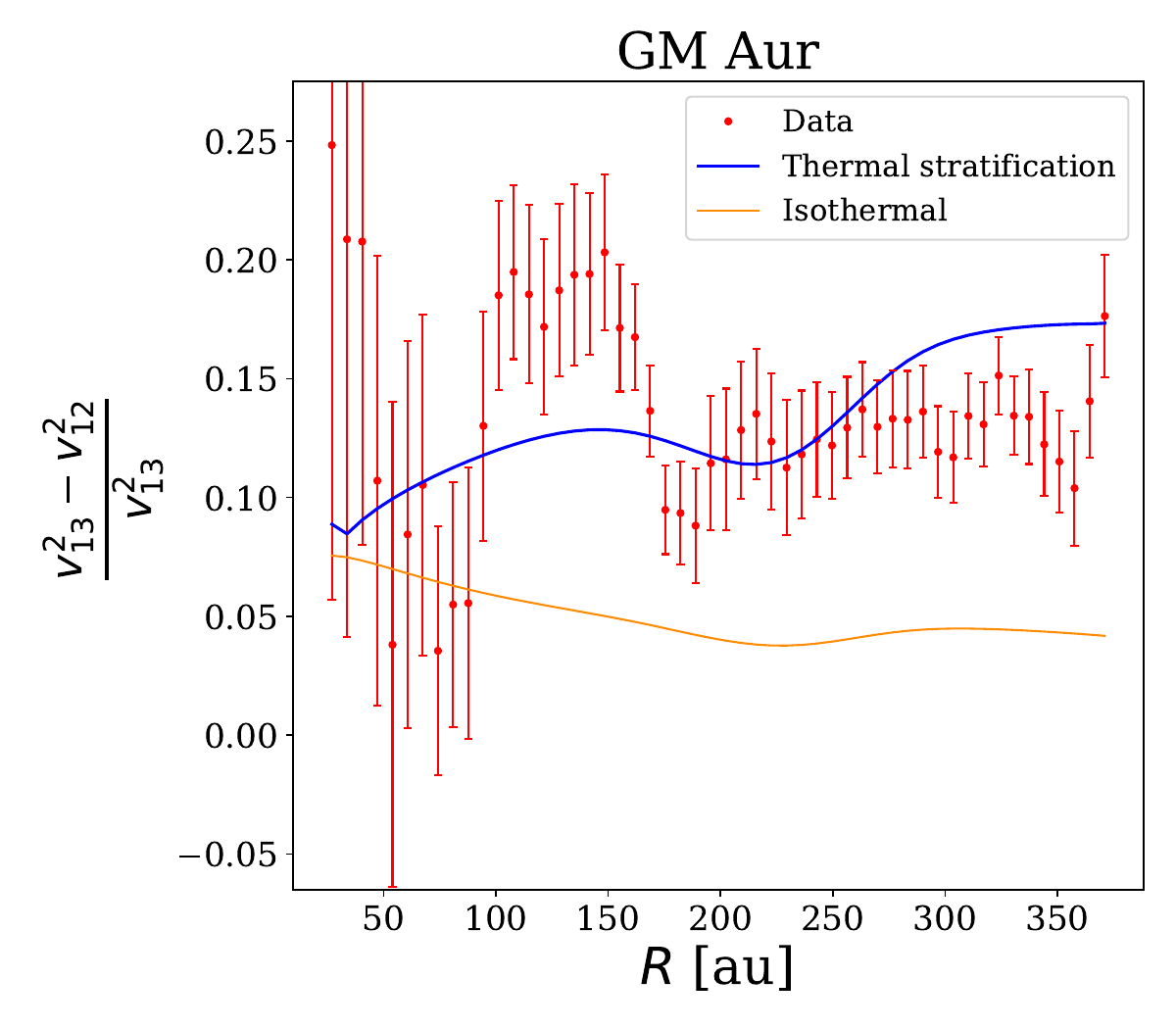}
    \includegraphics[scale=0.435]{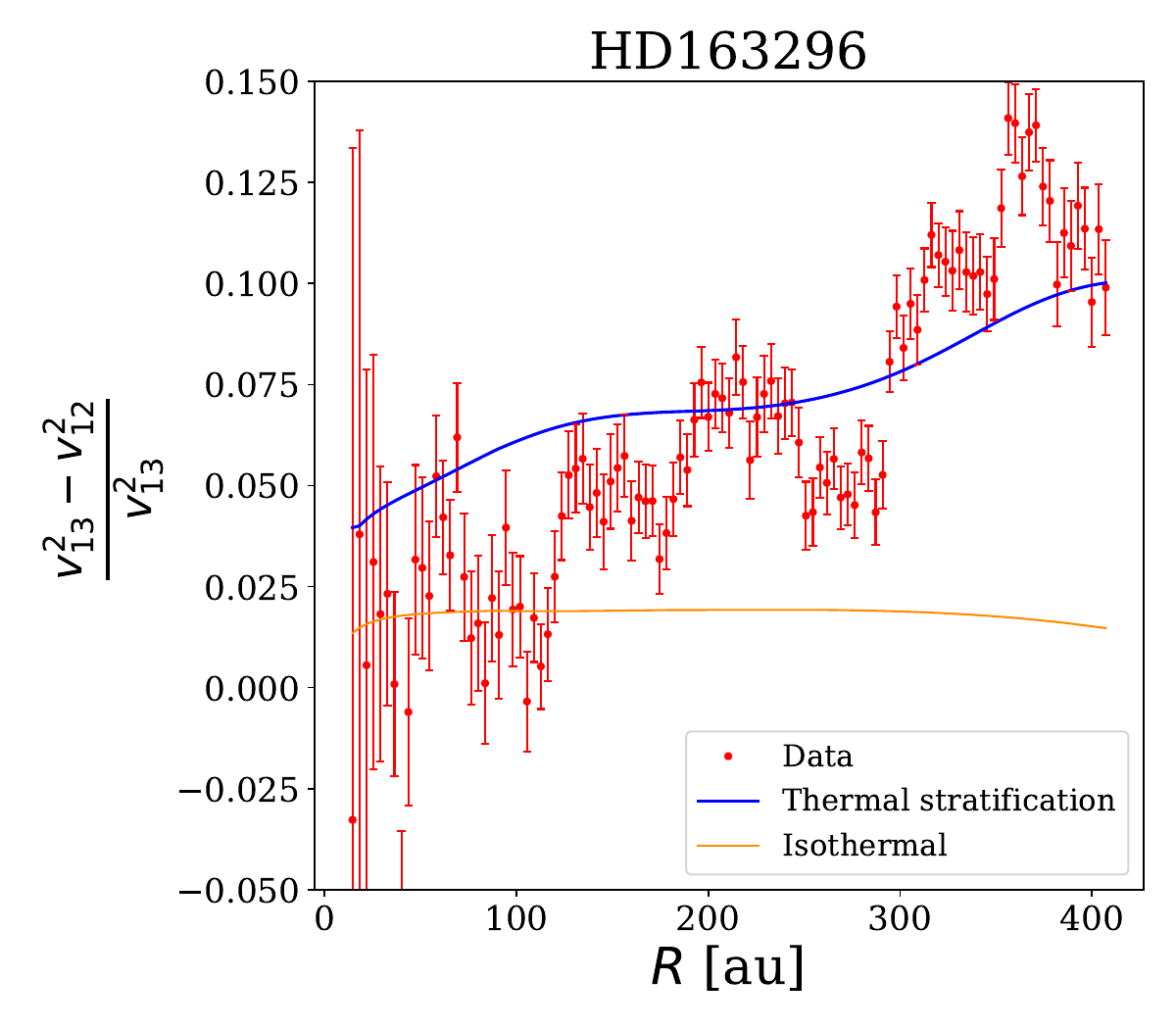}
        \includegraphics[scale=0.435]{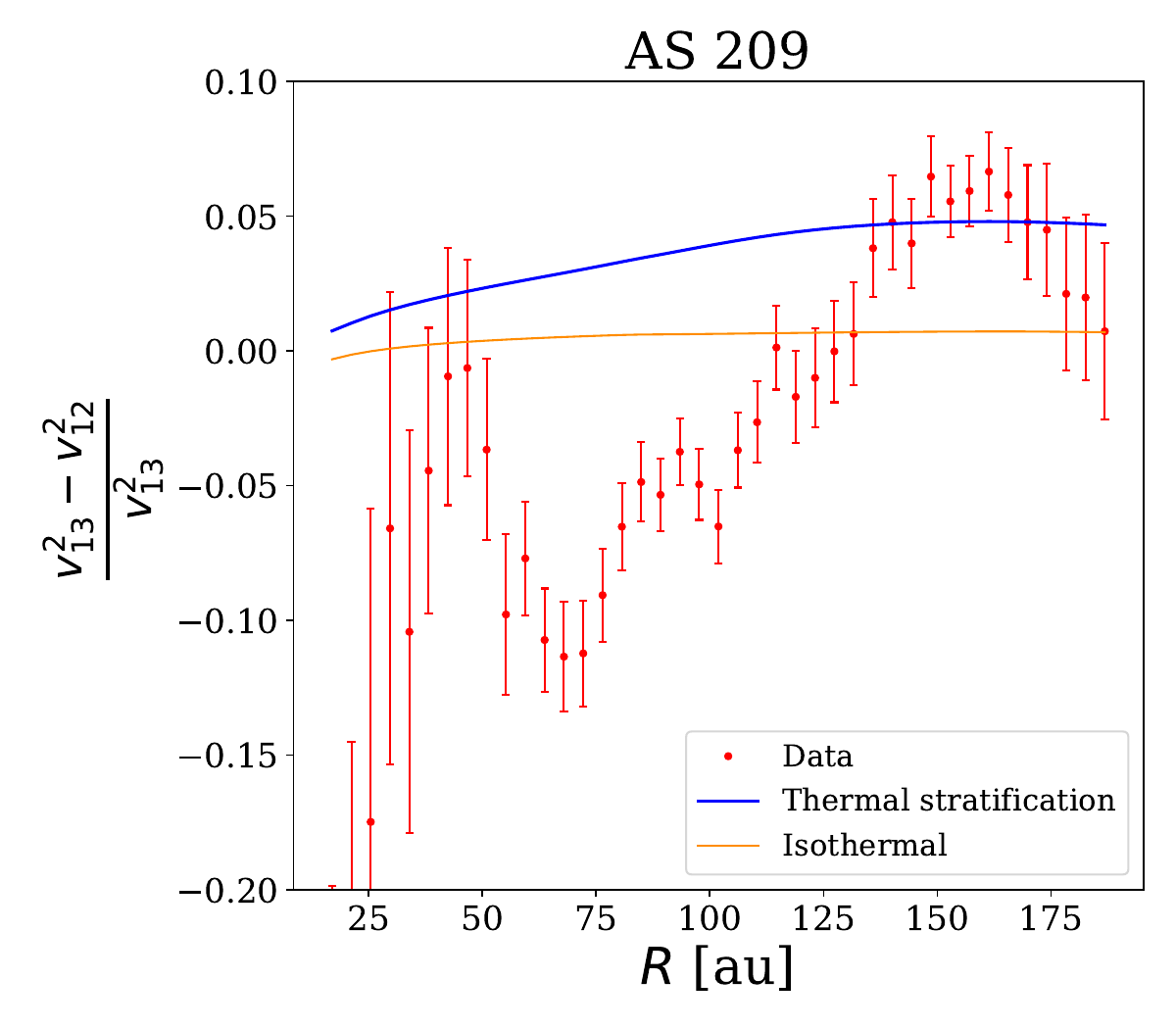}
      \caption{Relative difference between the squares of observed $^{12}$CO and $^{13}$CO rotation curves predicted by the thermally stratified model (blue line), the isothermal model (orange line), and the data (red dots). Except for AS 209, where this quantity is negative in the inner part, it is clearly visible that  data are well reproduced by the stratified model. Indeed, the difference of speed between the two curves cannot be explained just in terms of different height.}
    \label{diff_plot}    
\end{figure*}

\begin{figure*}[h]
    \centering
    \includegraphics[scale=0.45]{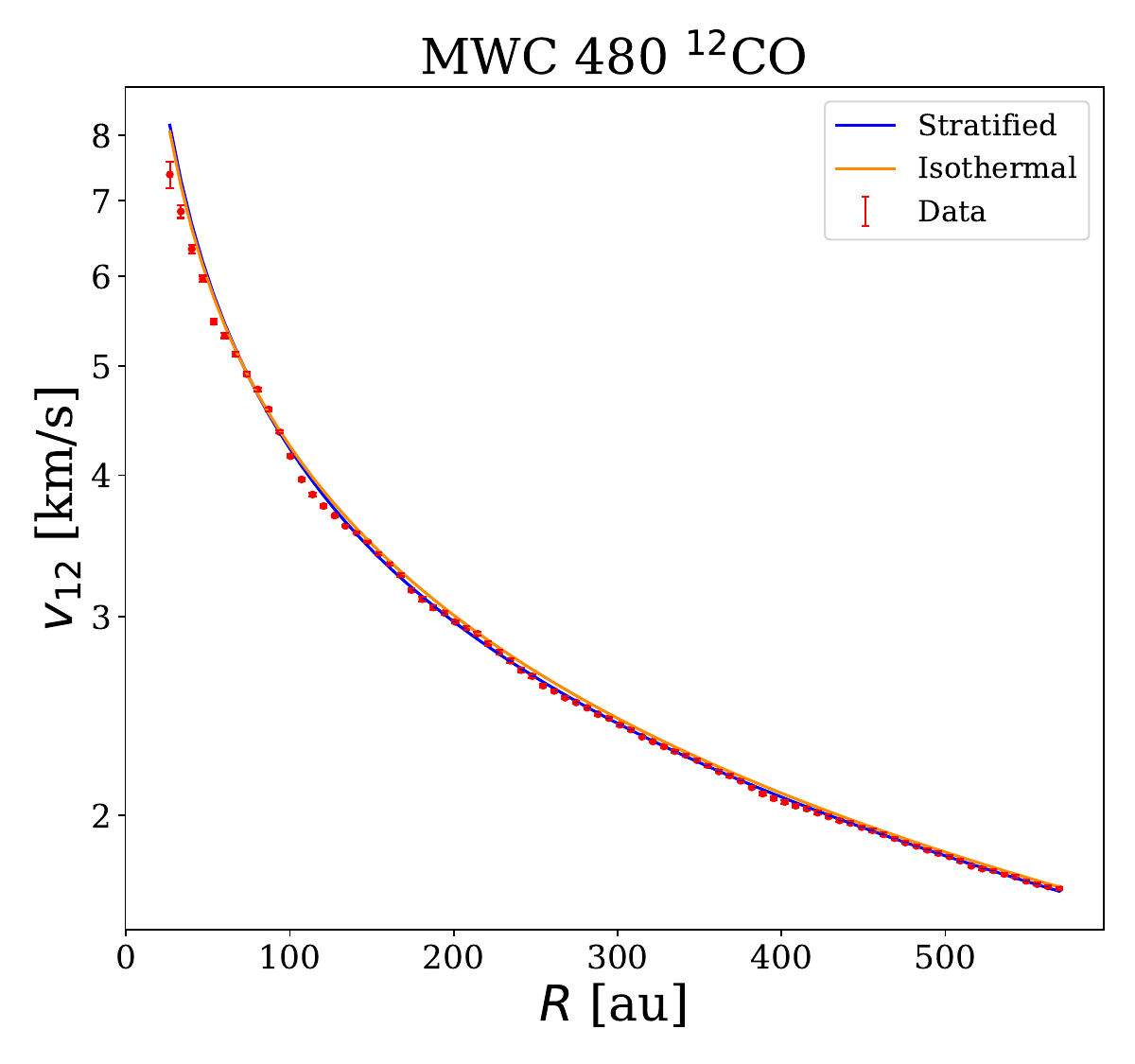}
    \includegraphics[scale=0.45]{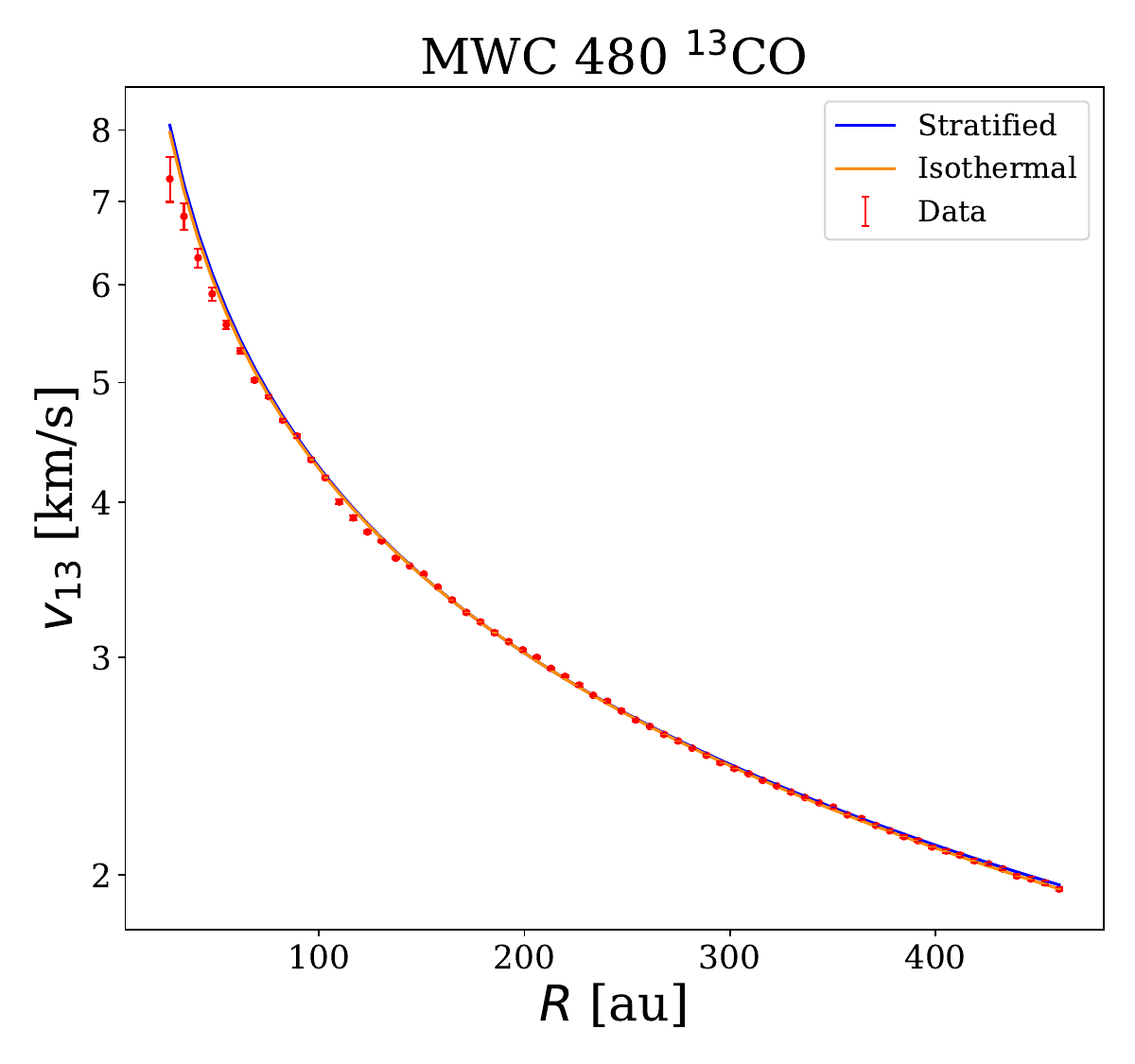}
      \caption{Left panel: Rotation curve of MWC 480 obtained from $^{12}$CO data (red points), along with our best-fitting curve for the stratified model (blue line) and the isothermal one (orange line). Right panel: same for the $^{13}$CO data.}
    \label{MWC plot}
\end{figure*}

\begin{figure*}[h]
    \centering
    \includegraphics[scale=0.45]{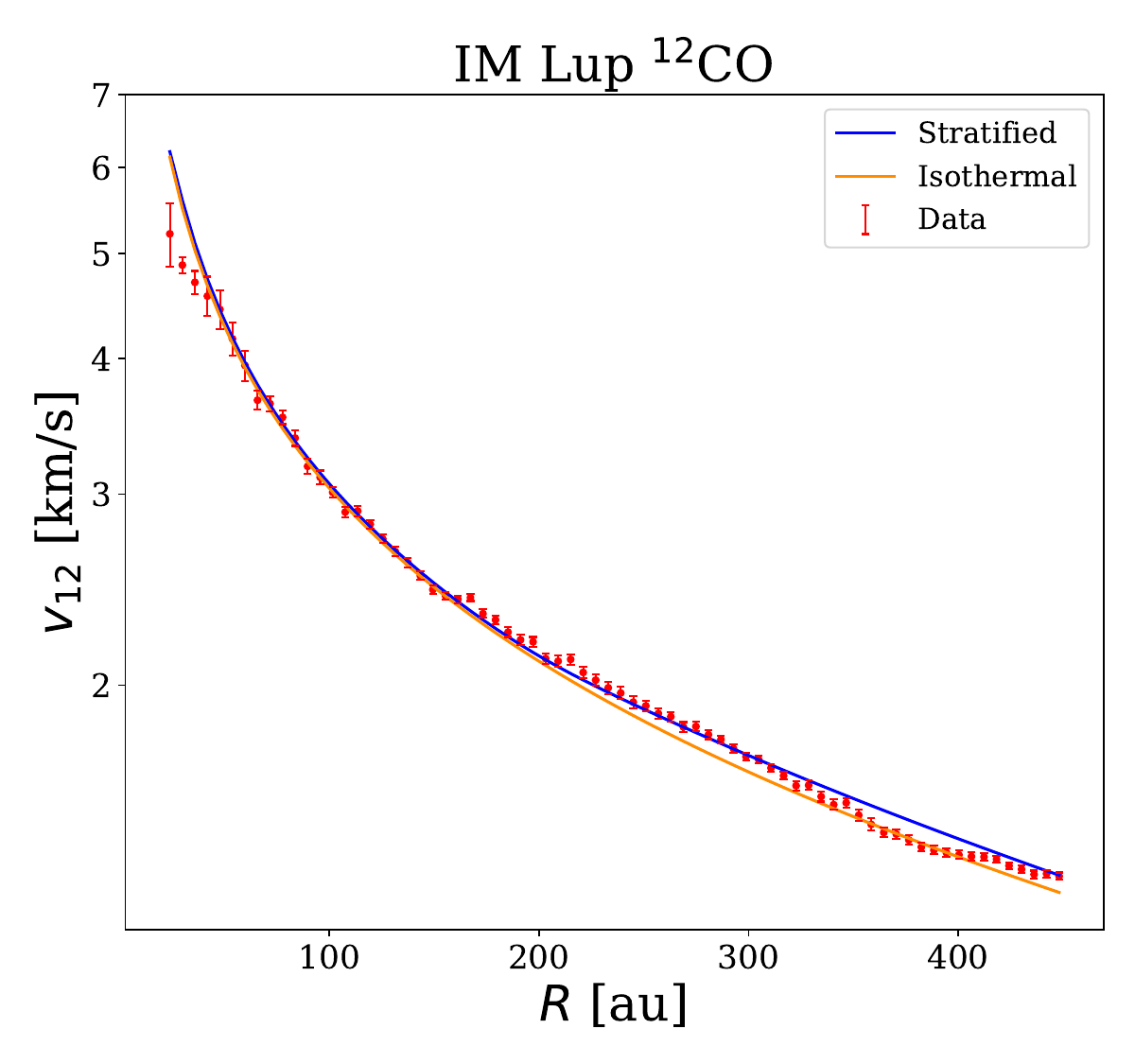}
    \includegraphics[scale=0.45]{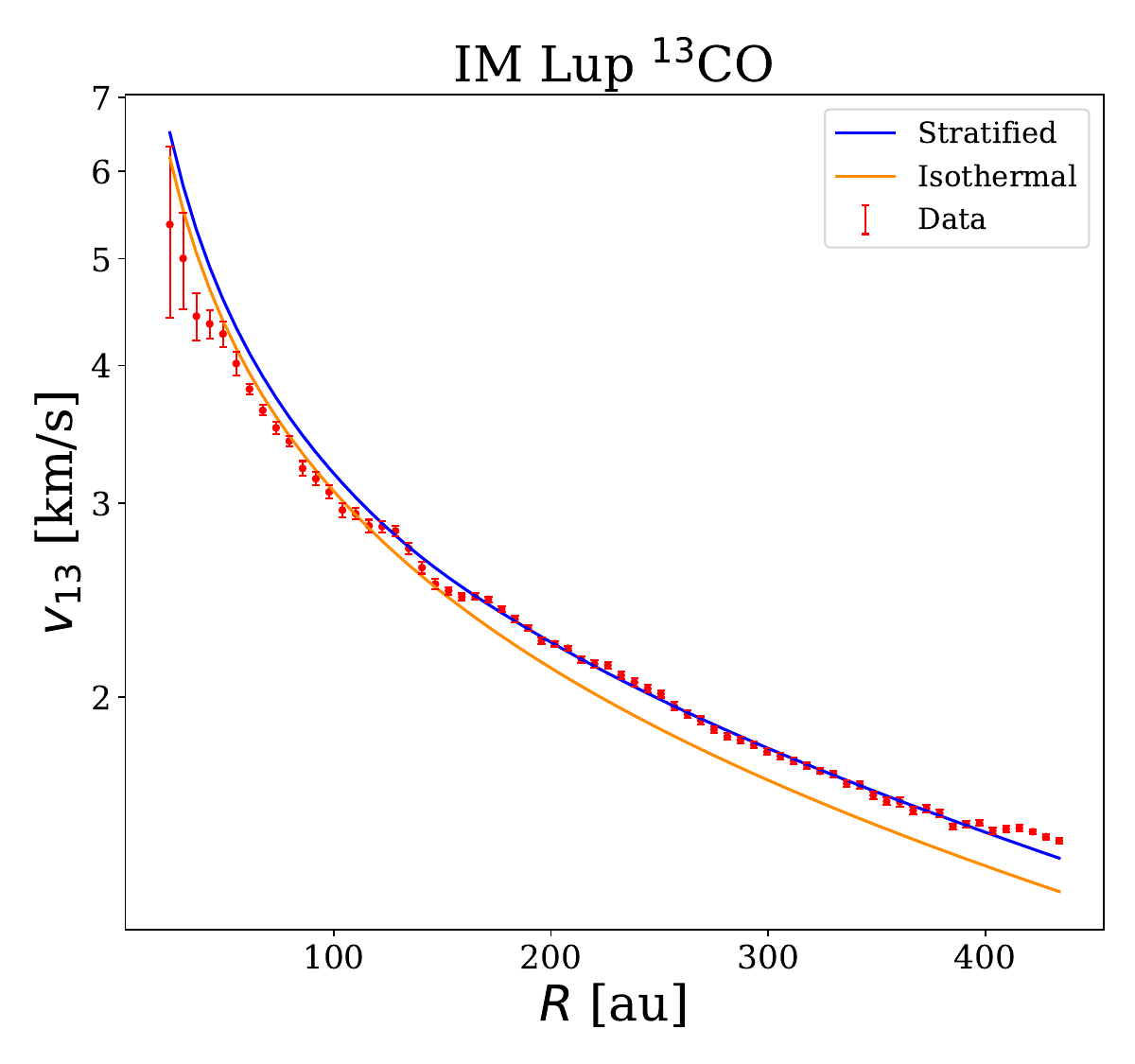}
      \caption{Same as Figure \ref{MWC plot} but for IM Lup. }
    \label{IM plot}
\end{figure*}

\begin{figure*}[h]
    \centering
    \includegraphics[scale=0.45]{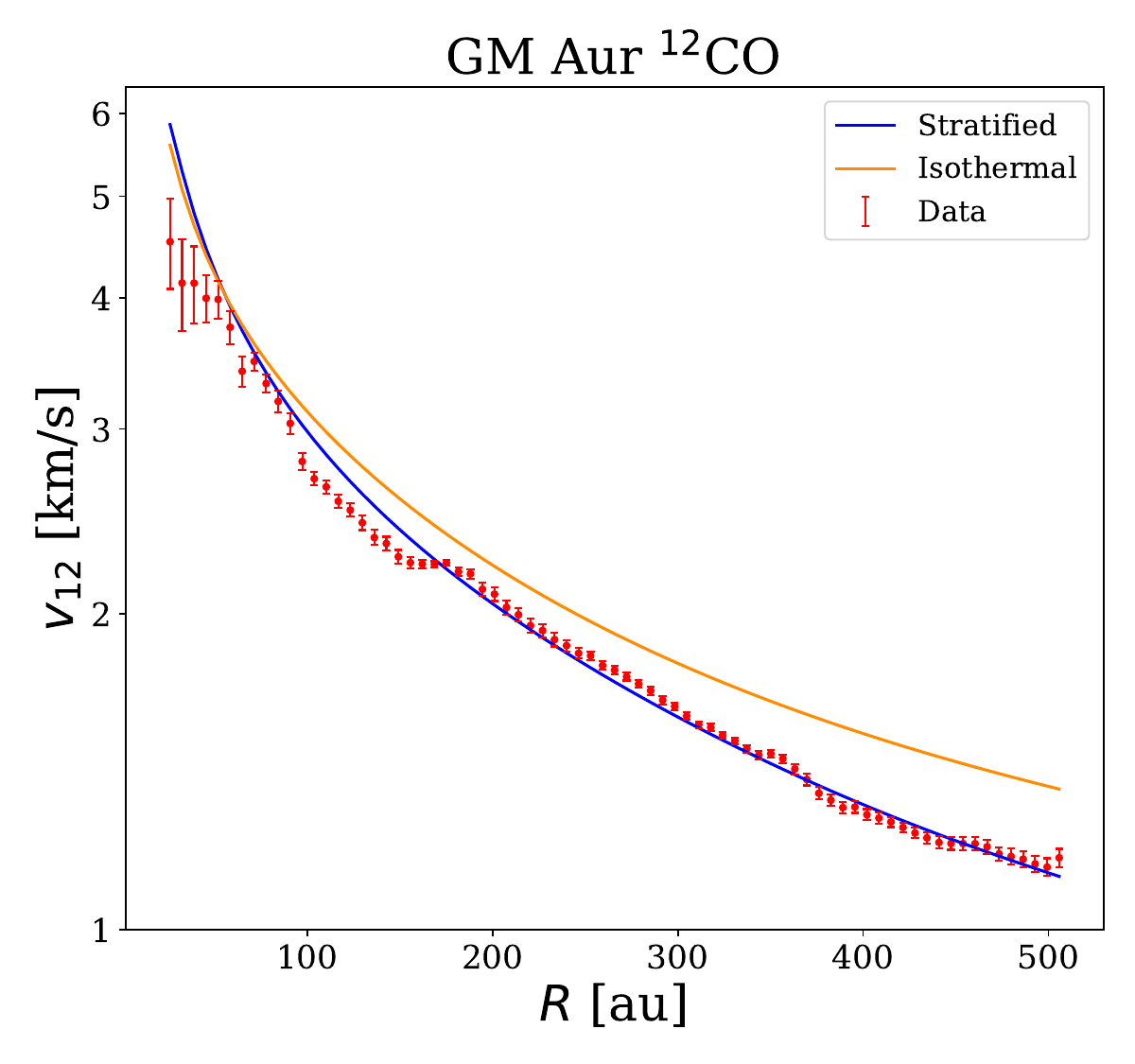}
    \includegraphics[scale=0.45]{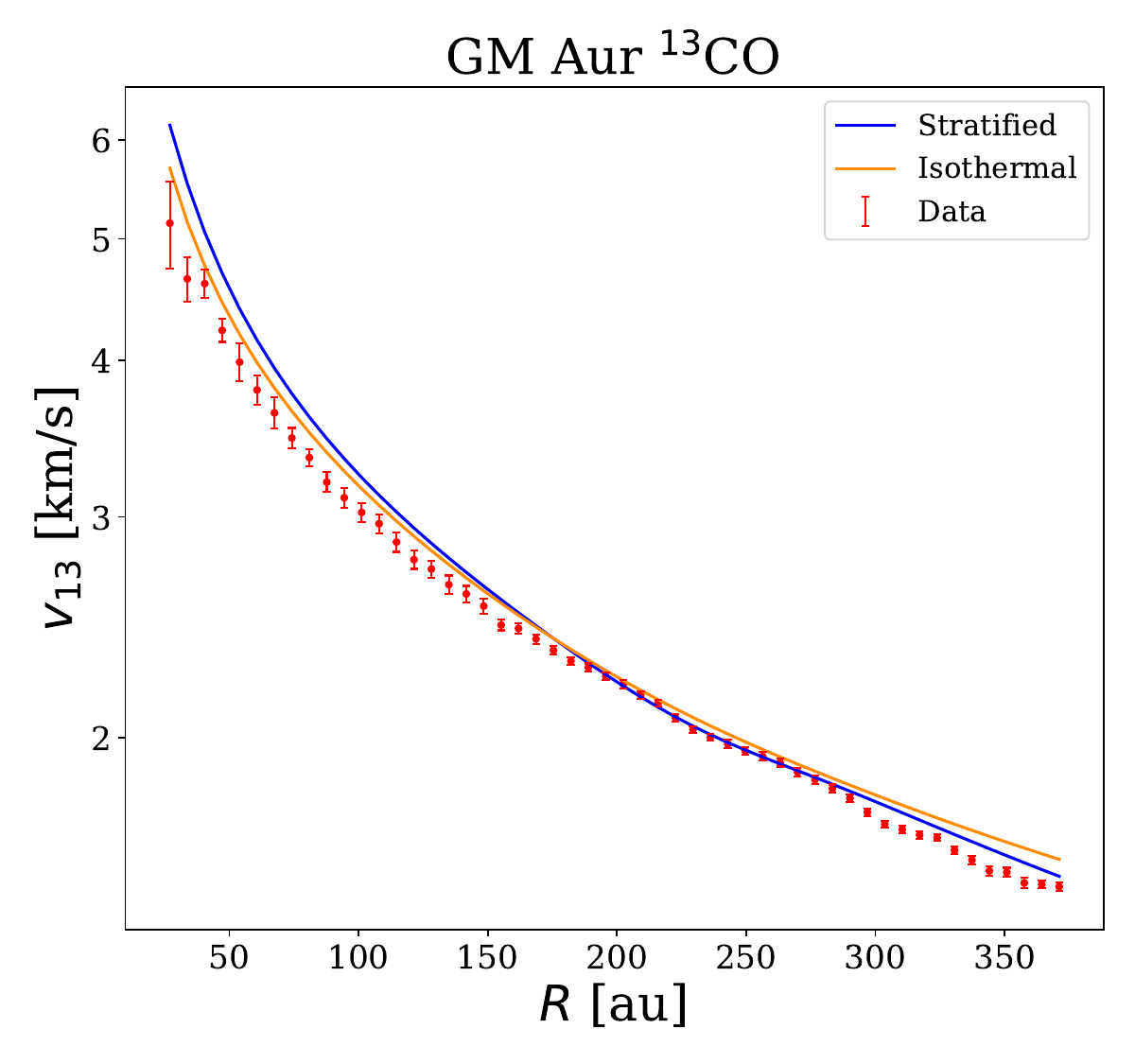}
      \caption{Same as Figure \ref{MWC plot} but for GM Aur.}
    \label{GM plot}
\end{figure*}

In this section, we describe how we applied the model to the entire sample of disks from the MAPS ALMA Large Program \citep{MAPSI}.
We performed our fits under the assumption of vertically isothermal or stratified disk to then compare the results. For the vertically isothermal model, the thermal structure is defined by the hydrostatic height of the disk at $R=100$au and the power law coefficient of the temperature profile, $q$. These parameters were taken from \cite{MAPSV}. As for the stratified model, \cite{MAPSIV} obtained the two-dimensional (2D) temperature structure of the MAPS disks, using the \citet{Dullemond} prescription given by Eq. \eqref{Dull}.  We note that the rotation curve traced by a specific molecule is defined by:
\begin{equation}
    v_\text{rot}^2(R)=v_\phi^2(R,z(R)),
\end{equation}
where $z(R)$ is the height of the emitting layer of the considered molecule.
For the emitting layer, we use:
\begin{equation}
    z(R) = z_0 \left(\frac{R}{100\text{au}}\right)^\psi \exp\left[-\left(\frac{R}{R_t}\right)^{q_t}\right],
\end{equation}
where the best-fit parameters have been obtained by \cite{izquierdo23}. All the parameters used are summarized in  Table \ref{table_params}. 

\begin{table}[H]
\caption{Velocity extraction method, orientation parameters, thermal parameters and emitting surfaces for $^{12}$CO and $^{13}$CO data of the MAPS disks. The orientation parameters and the emitting surfaces are taken from \cite{izquierdo23}, the thermal parameters for the isothermal model are taken from \cite{MAPSV}, for the stratified model from \cite{MAPSIV}.}\label{table_params}
\resizebox{\columnwidth}{!}{%
\begin{tabular}{llllll}
 & \textbf{MWC 480} & \textbf{IM Lup} & \textbf{GM Aur} & \textbf{HD 163296} & \textbf{AS 209}\\ \hline \\
 \textbf{Extraction} & & & & &\\ \\
 $^{12}$CO & Gauss & Dbell & Dbell & Dbell & Gauss \\ 
 $^{13}$CO & Gauss & Dbell & Dbell & Dbell & Gauss \\ \\
\textbf{Orientation} & & & & &\\ \\
$i$ [deg] &37.00 &47.50 &53.20 & 46.69& 35.00 \\
PA [deg] &328.15 &144.50  &53.98  & 312.75& 85.20\\ \\
\textbf{Isothermal} & & & & &\\ \\
$H_{100}$[au] & $10$ & $10$
&$7.5$ &$8.4$ &$6$\\
$q$
&$0.82$ & $0.66$ &$0.3$ &$0.84$ & $0.5$\\ \\
\textbf{Stratified} & & & & & \\ \\
$T_\text{mid}$[K]
&27 &25 &20 &24 &25\\
$T_\text{atm}$[K] &69 &36 &48 &63 &37 \\
$q$ &0.23 &0.02 &0.01 &0.18 &0.18 \\
$q_\text{atm}$ &0.7 &-0.03 &0.55 &0.61 &0.59 \\
$\zeta_0$[au] &7 &3 &13 &9 &5 \\
$\alpha$ &2.78 &4.91 &2.57 &3.01 &3.31\\
$\beta$ &-0.05 &2.07 &0.54  &0.42 &0.02\\ \\
\textbf{ $^{12}$CO Surface} & & & & & \\ \\
$z_0$[au] & 17.04&34.13 &32.00 &27.14 &16.47\\
$\psi$ &1.35 &0.99 &0.97  &1.07 &1.24\\
$R_t$[au] &579.43 &889.40 &729.91 
&534.00 &327.52\\
$q_t$ &1.63 &3.18 &3.22  &2.99 &3.01 \\ \\
 \textbf{ $^{13}$CO Surface} & & & & & \\ \\
 $z_0$[au] &11.52 &22.84 &18.21 &16.09 & 4.13\\
 $\psi$ &1.09 &1.27 &1.14  &1.12 &0.96\\
 $R_t$[au] &402.77 & 529.06&512.13 &392.75 &180.22 \\
 $q_t$ &1.87 &1.65 &2.73  &3.43 &3.59
 \end{tabular}%
}
\end{table}

The rotation curves (Fig. \ref{rcurves_1213}) can be obtained through different moment maps, according to the disk emission. 
We underline that rotation curve extraction 
we are only interested in measuring velocities from the frontside. Since three of the sources have strong 
contributions from the disk backside, we used a double-Bell decomposition to distinguish between these two 
components as introduced \citep{izquierdo22}. In this work, we have used an improved algorithm that 
performs this decomposition based on velocity priors obtained from the 
\textsc{discminer} models \citep{izquierdoprep}.

\subsection{Results}
We simultaneously fitted  the $^{12}$CO and $^{13}$CO data with both the isothermal and stratified model including the self gravitating contribution. The results are shown in Figs \ref{MWC plot}, \ref{IM plot}, \ref{GM plot}, \ref{HD plot}, and \ref{AS plot}, and the best-fitting parameters  reported in  Table \ref{fits} and Appendix \ref{app_corner}. 

To quantify the importance of thermal stratification, we computed the relative difference between the squares of $^{12}$CO and $^{13}$CO rotation curves, as shown in Fig. \ref{diff_plot}. According to the vertical isothermal model, this quantity is:
\begin{equation}
    \left(v_{13}^2 -v_{12}^2\right)_\text{iso} = v_\text{k}^2 q \frac{ \sqrt{1+z_{12}^2/R^2} - \sqrt{1+z_{13}^2/R^2} }{\sqrt{\left(1+z_{13}^2/R^2\right)  \left( 1+z_{12}^2/R^2 \right)}},
\end{equation}
which solely depends on the different height of the tracer, since it is assumed that the temperature does not change vertically. As for the stratified model, the expression is more complex, since it involves the evaluation of the term given by Eq.\eqref{fg} at different heights. In this case, we expect to observe larger differences between the velocity of the two isotopologues, since there is an additional shift caused by the different emission temperature. To determine the importance of vertical stratification, we quantified the maximum value of the velocity shift between $^{12}$CO and $^{13}$CO, which can be predicted in the isothermal case:
\begin{equation}\label{shift}
    \frac{\left(v_{13}^2 -v_{12}^2\right)_\text{iso}}{v_\text{k}^2}\approx q\frac{\Delta z^2}{2R^2}<5\%,
\end{equation}
where we assumed that typically $z/R<0.5$. Hence, if the quantity $(v_{13}^2 -v_{12}^2)/v_\text{k}^2$ is higher than $5\%$, the system cannot be described by an isothermal model, while it is likely that vertical stratification plays a significant role. It is important to note that the Eq. \eqref{shift} depends on the star mass through $v_k^2$. 
We normalized the squared differences of the velocities by the square of the velocity for $^{13}$CO since this quantity is independent of the stellar mass. Figure \ref{diff_plot} shows this quantity for the studied systems.

In the following, we will present the results of each disk, along with a discussion of the importance of thermal stratification. To compare the results, we performed our fits for both the vertically isothermal and stratified case. In addition, we computed the dust mass from millimetric emission at $283\text{GHz}$, using \citep{hilde84}:
\begin{equation}\label{dustmass}
    M_\text{dust} = \frac{d^2 F_\nu}{\kappa_\nu B_\nu(T)},
\end{equation}
where $d$ is the distance, $F_\nu$ is the flux density in Jy, $\kappa_\nu = 2.3 (\nu/230\text{GHz})^{0.4} \text{cm}^2\text{g}^{-1}$ is the dust opacity, and $B_\nu$ is the blackbody spectrum. In our analysis, we assumed $T=20$K and $\nu=283$GHz, while the flux densities were extracted from MAPS data. We recall that this equation implies that dust emission is optically thin. The results are reported in  Table \ref{gtdratio}.

%Another interesting source to test is IM Lup, studied in \cite{LodatoLongarini}. In that work, the authors were able to obtain a consistent model of both the isotopologues, and the rotation curves did not show any sign of thermal stratification. The results of the best-fit model were $M_\star = 1\text{M}_\odot$, $M_d = 0.1\text{M}_\odot$ and $R_c = 88\text{au}$. We evaluated the quantity \ref{shift} for IM Lup, and it is lower than $5\%$. Additionally, we plotted the difference between $^{12}$CO and $^{13}$CO rotation curves of the data, compared with the best-fit model of \cite{LodatoLongarini} in figure \eqref{imlup_strat}. The isothermal model well reproduces the data, and hence we decide not to include this source in our study. 

\subsection{MWC 480}
MWC 480 is a $\sim7$Myr Herbig Ae star located in the Taurus-Aurigae star-forming region at a distance of $d=162$pc \citep{montesinos09}. 
%Its disk was firstly targeted for continuum millimetric observation by \citet{pietu06} and afterwards with a resolution of $0.1^{\prime\prime}$ by \cite{long18}.  The dust continuum emission shows two dust rings and low emission out of 200au. Recently, \cite{teague21} and \cite{izquierdo23} showed kinematic evidence of an embedded protoplanet at a distance of $\sim 250$au, with associated vertical flows and buoyancy spirals. 
The most recent value of the stellar mass has been derived dynamically by \cite{izquierdo23} to be $M_\star = 1.97\text{M}_\odot$. \cite{MAPSV} through 2D thermochemical models computed disk mass and scale radius of the MAPS disks. For MWC 480, these values are $M_d = 0.16\text{M}_\odot$ and $R_c = 200$au. 

By inspecting the $^{12}$CO and $^{13}$CO rotation curves (Fig. \ref{rcurves_1213}), there is no evident sign of thermal stratification, since the two curves do not differ significantly. Figure \ref{MWC plot} shows that the two models are nearly indistinguishable, but in Fig. \ref{diff_plot} we see that the stratified model is better at reproducing the data.
When we assume an isothermal model, we obtain $M_\star = 1.969\pm0.002\text{M}_\odot$, $M_d = 0.201\pm0.002\text{M}_\odot$ and $R_c= 80\pm1$au, while for the stratified model, it is $M_\star = 2.027\pm0.002\text{M}_\odot$, $M_d = 0.150 \pm0.002\text{M}_\odot$ and $R_c = 128\pm1\text{au}$. The disk mass obtained with the stratified model is in agreement with the literature value \citep{MAPSV}. Since the reduced chi-squared $\chi^2_\text{red}$ is smaller in the stratified case (see  Table \ref{fits}), we adopted it as the best-fit model.

%A comparison between the two models is shown in figure \ref{MWC plot}, and they are nearly indistinguishable.Since the reduced chi-squared $\chi^2_\text{red}$ is smaller in the stratified case (see table \ref{fits}), we adopt it as the best-fit model, even though there are no evident differences between the two models. Although not significant, the kinematic signatures generated by thermal stratification are shown in figure \ref{diff_plot}, and the stratified model reproduces them very well. 

\subsection{IM Lup}

IM Lup is a young pre-main sequence star ($\sim 1$Myr) located in the Lupus star-forming region at a distance of 158pc \citep{Gaia18}. The dynamical stellar mass is estimated to be $1.1\text{M}_\odot$ \citep{teague21}, and it hosts an unusually large disk, extending out to $\approx 300$ au in the dust continuum and out to $\approx 1000$ au in the gas \citep{Cleeves16}. 
%This system is part of the DSHARP sample \citep{andrews18} and the 
The dust continuum emission shows clear evidence of a spiral morphology, which may be triggered by gravitational instability \citep{huang18}. \cite{Cleeves16} first estimated the disk mass from mm visibilities and found a massive disk of $0.2\text{M}_\odot$. \cite{Verrios22} claimed that the spiral structure of IM Lup could be generated by an embedded protoplanet. They performed numerical SPH simulations of planet-disk interaction and then post-processed them  to compare their results with CO, dust, and scattered light emission. Interestingly, a high disk mass $(\sim 0.1\text{M}_\odot)$ is required to match the scattered light image, so that the sub-micron sized grains could remain well coupled in the top layers of the disk. \cite{Cleeves16} first estimated the disk scale radius $R_c = 100$au by comparing SED to a simple tapered power-law density profile. Afterwards, \cite{Pinte18} analyzed CO data and found that a tapered power law density profile with $R_c = 284$au more optimally reproduces the data. They also analyzed the rotation curve of the disk and found that while the inner disk is in good agreement with Keplerian rotation around a $1\pm 0.1\mathrm{M}_{\sun}$ star, both the $^{12}$CO and the $^{13}$CO rotation curves become sub-Keplerian in the outer disk. The authors attributed this effect to the pressure gradient. \cite{LodatoLongarini} analyzed $^{12}$CO and $^{13}$CO rotation curves and fitted for star mass, disk mass and scale radius with an isothermal model. {In particular, the authors found that for the rotation curves extracted with \textsc{eddy} the best-fit are $M_\star = 1.012\pm0.003 M_\odot, M_\text{d}=0.096\pm0.003M_\odot, R_c=89\pm1$au and \textsc{discminer} are $M_\star = 1.02\pm0.02, M_\text{d}=0.10\pm0.01M_\odot, R_c=66\pm1$au.} We underline that in this work the rotation curves have been obtained again, and they are different from those of \cite{LodatoLongarini}. This is also true for the case of GM Aur. 
%They found that the best-fit value for the disk mass and scale radius are $M_\text{disk}\sim0.1-0.2\text{M}_\odot$ and $R_c\sim65-85$au, depending on the extraction method for the rotation curves is used\footnote{In particular, the authors found that for the rotation curves extracted with \textsc{eddy} the best-fit are $M_\text{d}=0.096\pm0.003M_\odot, R_c=89\pm1$au and \textsc{discminer} are $M_\text{d}=0.10\pm0.01M_\odot, R_c=66\pm1$au.}. 

Figure \ref{IM plot} shows both the isothermal and stratified fit. While for $^{12}$CO, both  models describe the rotation curve well, for  $^{13}$CO the isothermal model fails, since the velocity shift is so high that it cannot be explained just in terms of emitting surface. This difference is clearly visible when considering the $\chi^2_\text{red}$, which for the stratified model is considerably smaller. The best-fit parameters for the isothermal model are $M_\star = 1.055\pm0.002\text{M}_\odot$, $M_d=0.200\pm0.003\text{M}_\odot$, and $R_c=55\pm1$au; while for the stratified model are $M_\star = 1.1994\pm0.002\text{M}_\odot$, $M_d=0.106\pm0.002\text{M}_\odot$, and $R_c=115\pm1$au. The effects of thermal stratification are visible in Fig. \ref{diff_plot}. At $R\sim 250$au, the difference in the data between $^{12}$CO and $^{13}$ CO is on the order of $\sim 10\%$ and it significantly increases in the outer part. There, neither the stratified model is able to explain that difference. \cite{izquierdo23} pointed out that the emission from the outer disk is so diffuse that the retrieval of the emitting surface, as well as the velocity extraction, needs to be taken with care. This is possibly an effect of external photoevaporation. Indeed, despite the very weak external radiation field irradiating IM Lup, \cite{haworth17} showed that the disk is sufficiently large that the outer part, which is weakly gravitationally bound, can undergo photoevaporation.

\subsection{GM Aur}
GM Aur is a T-Tauri star in the Taurus-Auriga star-forming region that hosts a transition disk. The stellar mass has been estimated dynamically to be $M_\star = 1.1 \text{M}_\odot$ by \cite{teague21}, in agreement with previous measurements \citep{macias18}. Its CO morphology is very complex, showing spiral arms, tails, and interactions with the environments \citep{huang21}. From thermochemical models of MAPS data, \cite{Schwarz21} obtained a disk mass of $M_d=0.2\text{M}_\odot$ and a scale radius of $R_c=111$au, making GM Aur a possibly gravitationally unstable disk. \citet{LodatoLongarini} performed a fitting procedure for the star mass, disk mass, and scale radius, using an isothermal model, finding that for GM Aur, the two CO lines provide inconsistent rotation curves, which cannot be attributed only to a difference in the height of the emitting layer. In addition, the authors provided a simple order-of-magnitude estimate of the expected velocity shift due to thermal stratification, concluding that the difference between the two rotation curves could not be explained by this effect. They drew this conclusion by taking into account the different temperature of the two molecules at their emission height, $z_i(R)$, given by \cite{MAPSIV}. However, as shown in Appendix \ref{calc}, in the azimuthal velocity, it is not only essential to know the temperature at $(R,z)$, but also its radial and vertical gradient at that location. 

By analyzing the rotation curves of the two CO isotopolgues (Fig. \ref{rcurves_1213}), a systematic shift between $^{12}$CO and $^{13}$CO curves is clearly visible, which may possibly be attributed to thermal stratification. When we perform the fitting with the isothermal model, we obtain (as the best-fit parameters) $M_\star = 0.872\pm0.003\text{M}_\odot$, $M_d = 0.312 \pm0.003\text{M}_\odot$, and $R_c= 56\pm1$au. This is in agreement with \cite{LodatoLongarini} and leads to a high $\chi^2_\text{red}$(see  Table \ref{fits}). As a matter of fact, Fig. \ref{GM plot} shows that an isothermal model is not able to reproduce both $^{12}$CO and $^{13}$CO rotation curves. Conversely, when thermal stratification is taken into account, the two rotation curves are compatible and are in agreement with data, especially for $R>180$au.
In this case, the best-fit value for the star mass is $M_\star = 1.128\pm 0.002 \text{M}_\odot$, which is in line with the literature values \citep{teague21,macias18}. As for the disk mass, the best-fit value is $M_d = 0.118\pm0.002 \text{M}_\odot$.
Finally, the best-fit value for the scale radius is $R_c = 96\pm 1 \text{au}$, almost twice the value obtained with the isothermal model and in good agreement with \cite{Schwarz21}. A stratified model reproduces the difference between $^{12}$CO and $^{13}$CO rotation curves  very well, as shown in Fig. \ref{diff_plot}, which leads to a significant decrease in the $\chi_\text{red}^2$ value.
%In figure \ref{difference rot curve} we show that the isothermal model does not explain the rotational difference between $^{12}$CO and $^{13}$CO, while the stratified model does. For GM Aur, the thermal stratification is able to recover not only the value of the difference, but also its radial morphology, showing the double peaked profile.

\begin{figure*}[h!tbp]
    \centering
    \includegraphics[scale=0.45]{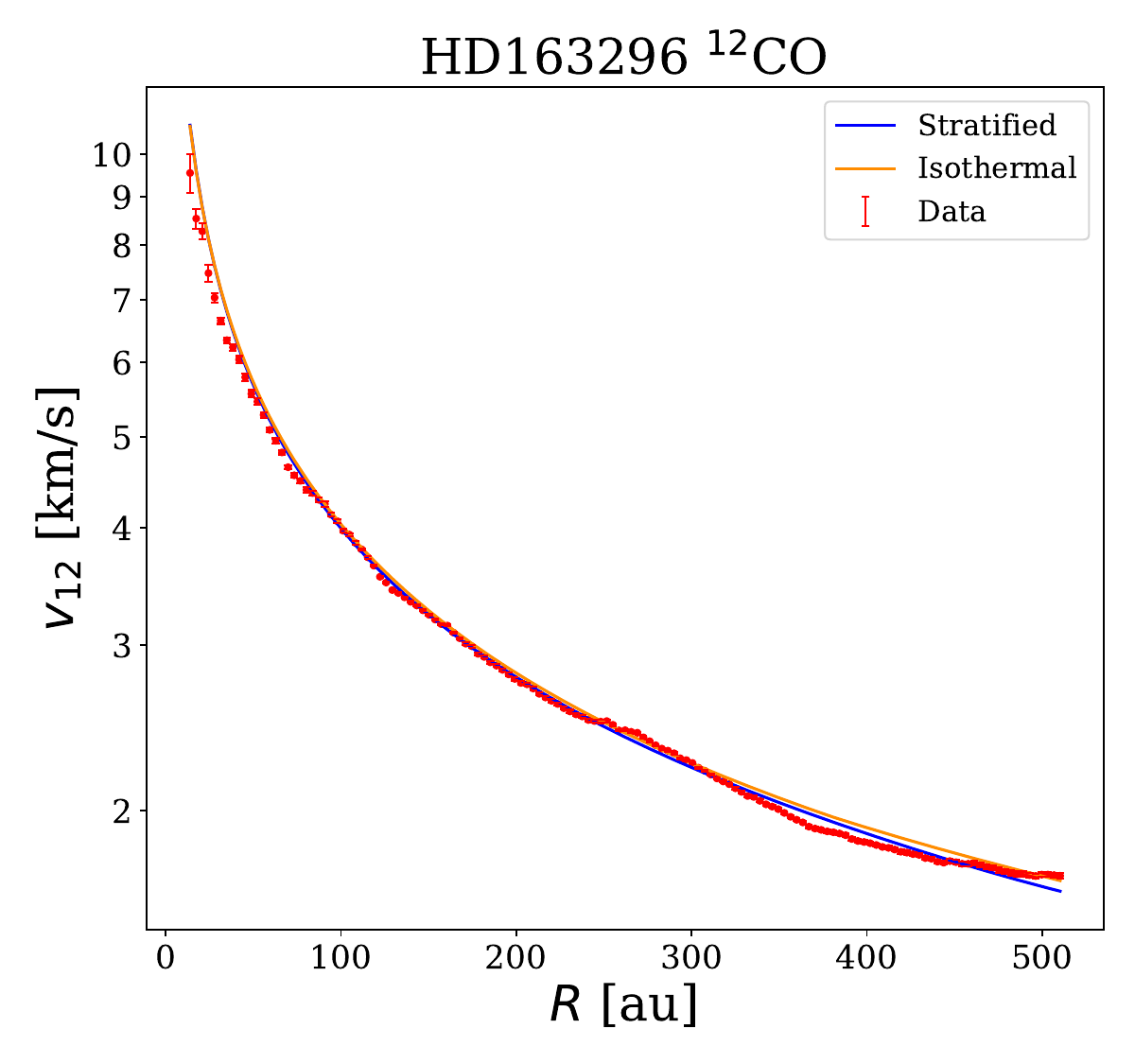}
    \includegraphics[scale=0.45]{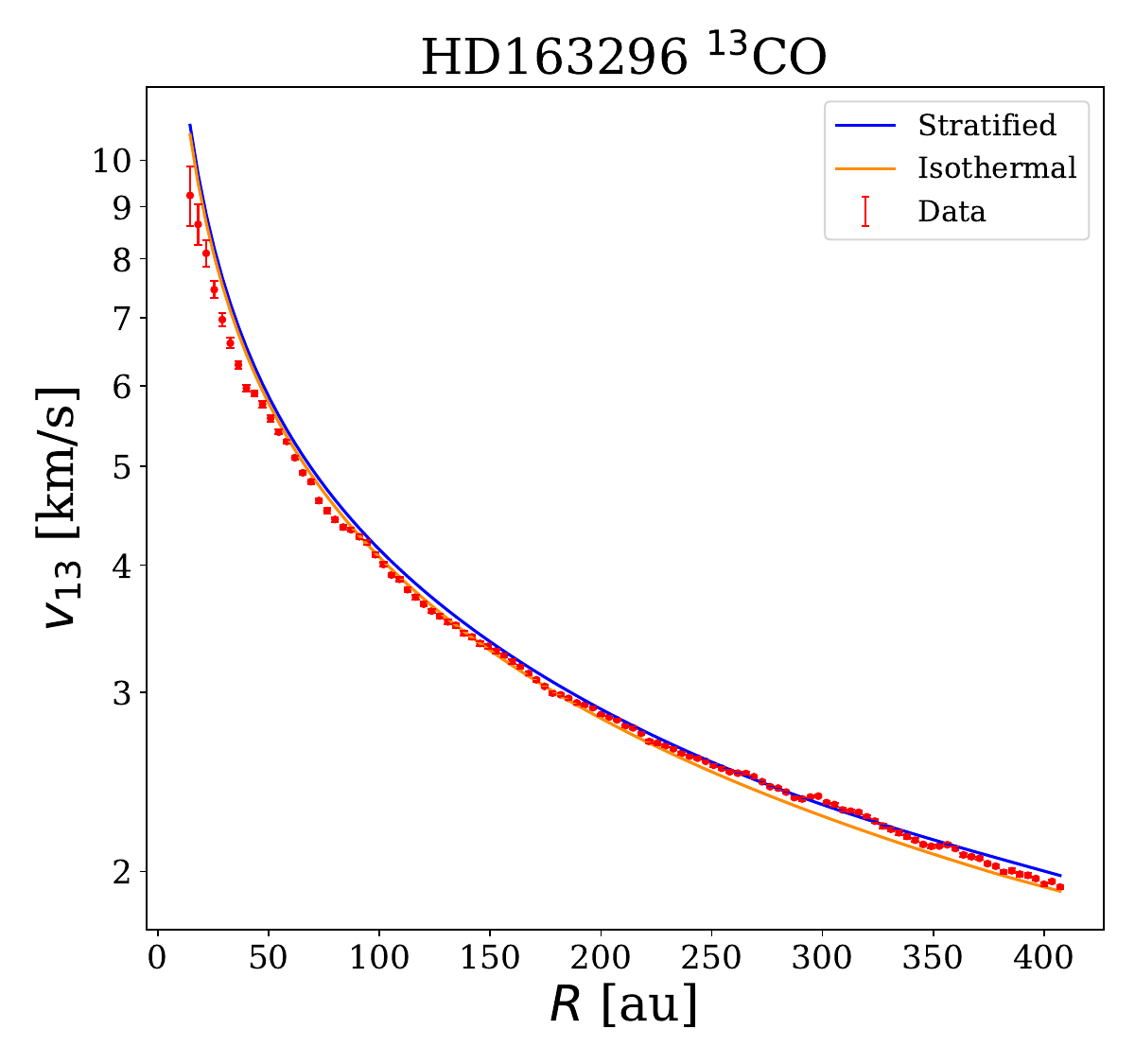}
      \caption{Same as Fig. \ref{MWC plot}, but for HD 163296.}
    \label{HD plot}
\end{figure*}

\begin{figure*}[h! tbp]
    \centering
    \includegraphics[scale=0.45]{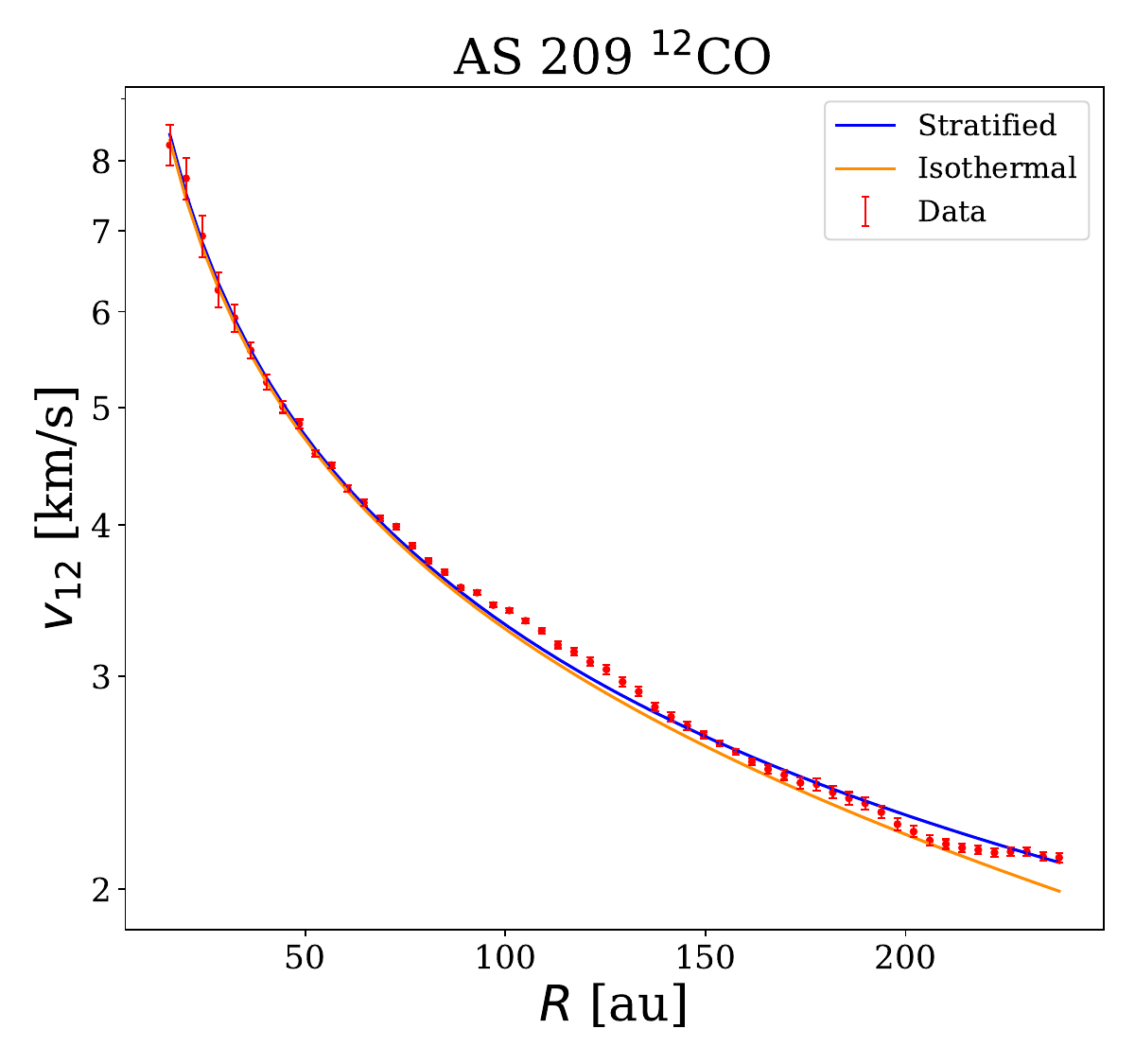}
    \includegraphics[scale=0.45]{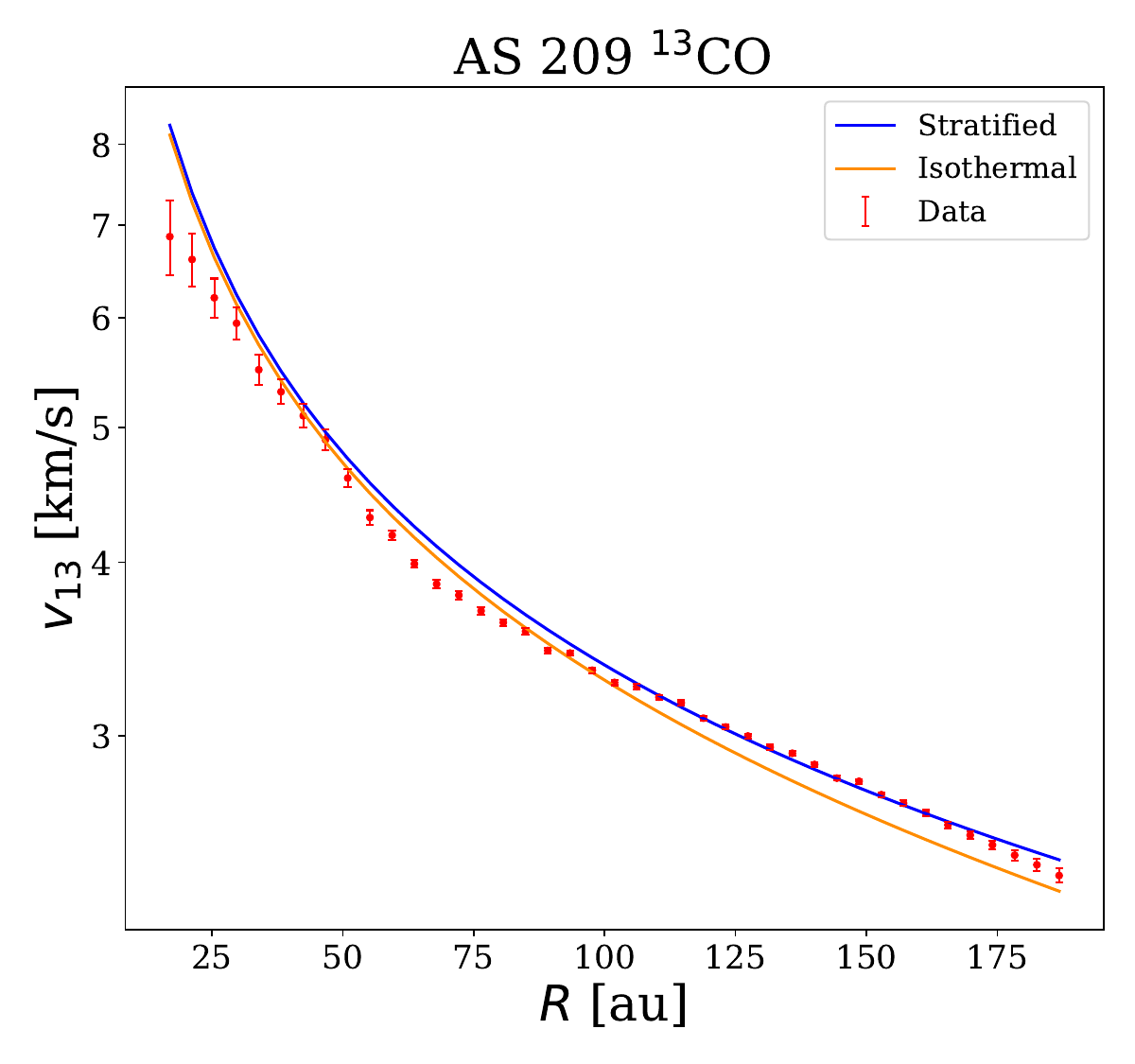}
      \caption{Same as Fig. \ref{MWC plot}, but for AS 209.}
    \label{AS plot}
\end{figure*} 

\subsection{HD 163296}
HD 163296 is one of the most well-studied Herbig Ae star system at millimeter wavelengths thanks to its relative close distance ($d=101$pc) and bright disk. The disk presents several features that suggest ongoing planet formation, such as dust rings, deviations from Keplerian velocities due to gas pressure variations, "kinks" in the CO emission, and meridional flows \citep{isella16,isella18,Pinte18,teague18a,pinte2022,izquierdo22, calcino22, izquierdo23}. \\
This system has also been extensively studied because there is considerable  evidence to support a massive disk.  By modeling the dust lines, \cite{powell19} found that the disk mass is $M_d =0.21\text{M}_\odot$. As for the scale radius, \citet{Degreg} through radiative transfer modeling found that $R_c=125$au is the value that better reproduces dust and CO ALMA observations. \cite{guidi16} presented a multiwavelength ALMA and VLA study of the disk and, via a modeling of the visibilities, they  found that the best-fit value of the scale radius is $R_c=118$au, in agreement with \cite{Degreg}.

When we fit data with a vertical isothermal model, we obtain (as the best-fit parameters) $M_\star = 1.842\pm0.002\text{M}_\odot$, $M_d = 0.124\pm0.001\text{M}_\odot$ and $R_c= 38\pm1$au. While the star mass is realistic, the scale radius is unrealistically small compared to the gas emission extent on the order of $400$au \citep{MAPSIV}. Additionally, the isothermal model is not able to reproduce the difference between the rotation curves of the two CO isotopologues (see Fig. \ref{diff_plot}), resulting in a relatively poor fit with a large $\chi^2_\text{red}$. If we include the 2D thermal structure, the quality of the fit increases (see $\chi^2_\text{red}$ in  Table \ref{fits}). In this case, the best-fit for stellar mass and disk mass does not change significantly ($M_\star = 1.948\pm 0.002M_\odot, M_\text{d}=0.134\pm0.001M_\odot$), while the scale radius does shift to $R_c=91\pm 1\text{au}$. Comparing our result for the disk mass to the literature values, we observe that our fit gives a value that is roughly half. Figure \ref{HD plot} shows that both the isothermal and the stratified model aptly describe the rotation curve of $^{12}$CO and $^{13}$CO. However, the shift between them, presented in Fig. \ref{diff_plot}, is well recovered only by the stratified model, which offers only a partial explanation for the significant increase of the plotted quantity. 
The presence of pressure modulated substructures in the rotation curves \citep{izquierdo23} impacts the quality of the fit (clearly visible in Fig. \ref{diff_plot}). One possible solution would be to include them in the fitting model.

\subsection{AS 209}
AS 209 is a young T-Tauri star in the Ophiucus star forming region ($d\sim121$pc). 
%The disk is part of the DSHARP sample \citep{andrews18}, and it shows a high level of structure, with several nested rings \citep{andrews09,fedele18,andrews18,huang18}. Also the gas emission is complex, showing an annular gap at $R\sim 200$au between two bright rings \citep{law21a}. Inside this gap, \cite{bae22} found a circumplanetary disk candidate. In addition, from the same strong outward vertical motions have been observed in CO emission, possibly imputed to MHD effects \citep{galloway23}. 
The most recent stellar mass estimate is $M_\star =1.14\text{M}_\odot$ \citep{izquierdo23}. \cite{fedele18} gave an estimate for the scale radius $R_c = 80$au through the modeling of the mm visibilities . Afterwards, through thermochemical modeling, they found a dust mass of $M_\text{dust}=3.5\times10^{-4}\text{M}_\odot$; with a gas-to-dust ratio of 100, this translates into $M_d = 0.0035\text{M}_\odot$, which is in agreement with the recent value $M_d = 0.0045\text{M}_\odot$ of \cite{MAPSV}.
Interestingly, when inspecting the rotation curves of AS 209 (Fig. \ref{rcurves_1213}), the $^{13}$CO is slower compared to the $^{12}$CO, despite it being closer to the midplane. This trend is observed up to $\sim 125$au. 
A possible explanation for this is the compactness of the disk, which makes more difficult to extract a precise emitting surface due to beam smearing. Indeed, line centroids from pixels near the center of the disk are an averaged composition of multiple surrounding velocities because of  the limited resolution and the steepness of $v(r)$. Since AS209 is the smallest disk in the sample, it is more prone to be affected by this in the largest fraction of its total extent compared to the other sources.
When we fit with the isothermal model, we obtained, as the best-fit parameters, $M_\star = 1.272\pm0.003\text{M}_\odot$, $M_d = 0.042 \pm0.003\text{M}_\odot$, and $R_c= 45\pm1$au. When we fit with the stratified model, we obtain as the best-fit parameters $M_\star = 1.311\pm0.001\text{M}_\odot$, and $R_c= 126\pm2$au, while for the disk mass we report a $3-\sigma$ upper limit of  $M_d = 0.00025 \pm0.00025\text{M}_\odot$, since the best-fit parameter is compatible with zero. Both models are shown in Fig. \ref{AS plot}. As for $^{12}$CO, the two models behave in the same way, showing little difference in the outer edge. Conversely, for $^{13}$CO the isothermal model works better in the inner part, where $^{13}$CO is slower, while in the outer part the stratified model describes well the rotation curve. According to the $\chi^2_\text{red}$, the stratified model describes the data better (see  Table \ref{fits}).

\begin{table}\caption{Results of the fitting procedure and reduced chi-squared for the two different models:  isothermal and stratified.}\label{fits}
\resizebox{0.475\textwidth}{!}{
\begin{tabular}{lllll}
 & $M_\star$ [M$_\odot$] & $M_d$ [M$_\odot$] & $R_c$ [au] & $\chi^2_\text{red}$\\ \hline \\
 \textbf{MWC 480} & & & &  \\ \\
\textit{Isothermal} & 1.969 $\pm 0.002$ & 0.201$\pm 0.002$ & 80$\pm 1$ & 11.21 \\
\textit{Stratified} & 2.027$\pm 0.002$ & $0.150\pm 0.002$ & 128 $\pm 1$ & 6.14 \\ \\
\textbf{IM Lup} & & & &  \\ \\
\textit{Isothermal} & 1.055 $\pm 0.002$ & 0.200$\pm 0.003$ & 55$\pm 1$ & 35.68 \\
%\textit{Stratified SG} & 1.036$\pm 0.002$ & 0.033$\pm 0.003$ & 174 $\pm 2$ \\ \\
\textit{Stratified} & 1.194$\pm 0.002$ & $0.106\pm 0.002$ & 115 $\pm 1$ & 6.29 \\ \\
\textbf{GM Aur} & & & & \\ \\
\textit{Isothermal} & 0.872 $\pm 0.003$ & 0.312$\pm 0.003$ & 56$\pm 1$ & 90.84  \\
\textit{Stratified} & 1.128$\pm 0.002$ & 0.118$\pm 0.002$ & 96 $\pm 1$ &  8.48  \\ \\
%\textit{Stratified No SG} & 1.057$\pm 0.004$ &  & 211 $\pm 10$ &  8.58 \\ \\
\textbf{HD 163296} & & & &  \\ \\
\textit{Isothermal} & 1.842 $\pm 0.002$ & 0.124$\pm 0.001$ & 38$\pm 1$ & 29.60 \\
%\textit{Stratified SG} & 1.036$\pm 0.002$ & 0.033$\pm 0.003$ & 174 $\pm 2$ \\ \\
\textit{Stratified} & 1.948$\pm 0.002$ & $0.134\pm 0.001$ & 91 $\pm 1$ & 19.74 \\ \\
\textbf{AS 209} & & & &  \\ \\
\textit{Isothermal} & 1.272 $\pm 0.003$ & 0.042$\pm 0.003$ & 45$\pm 1$ & 25.13 \\
%\textit{Stratified SG} & 1.036$\pm 0.002$ & 0.033$\pm 0.003$ & 174 $\pm 2$ \\ \\
\textit{Stratified} & 1.311$\pm 0.001$ & $0.0002\pm 0.0002$ & 126 $\pm 2$ & 10.55 \\ \\
\end{tabular}
}
\end{table}

%According to eq. \eqref{dustmass}, the dust mass of AS 209 is $M_\text{dust} = 0.034\times 10^{-2}\text{M}_\odot$. According to \cite{Veronesiprep}, the minimum measurable disk mass is $5\%$ of the star mass. This condition allows us to give an upper limit for the gas-to-dust ratio: indeed, if the disk mass was exactly $5\%$ of stellar one ($0.065\text{M}_\odot$), the gas-to-dust ratio would be 192. Hence, the actual gas-to-dust ratio is lower than 192. 

\section{Discussion}
\label{S5}
\subsection{Thermal stratification in MAPS disks}
Table \ref{fits} presents a summary of the findings of this study, comparing the isothermal model with the stratified one. It is evident from the results that the reduced $\chi^2$ value consistently decreases when employing the stratified model. This indicates that the inclusion of thermal stratification provides a more effective way of describing the observed data. In this context, MWC 480 is particularly interesting. Despite the small kinematic signatures of thermal stratification, as depicted in Fig. \ref{diff_plot}, the quality of the stratified fit is higher and it yields more reliable values for star mass, disk mass, and scale radius. On the opposite side, GM Aur is the system that shows the strongest effects of thermal stratification, given that the $^{12}$CO and $^{13}$CO systematically shifted over all the radial extent of the disk. The introduction of thermal stratification is able to reconcile these differences, reducing the $\chi^2_\text{red}$ by an order of magnitude. The only case where the stratified model encounters challenges in accurately describing both curves is in AS 209. This system is peculiar because  the compactness of the disk influences the extraction of emission surfaces. Consequently, contrary to what expected, we observe that the $^{13}$CO rotates slower than the $^{12}$CO in the inner part. Despite that, the $\chi^2_\text{red}$ is smaller when thermal stratification is taken into account.

\subsection{Disk masses}
In this paragraph, we aim to contextualize our work within the broader framework of disk mass estimation. 

One solid tracer of the disk mass is the carbon dioxide HD, which is a good tracer of the disk gas because it follows the distribution of molecular hydrogen and its emission is sensitive to the total mass. The first detection of HD emission in a protoplanetary disk comes from \cite{bergin13} for TW Hya. Afterwards, the detection of HD $J=1-0$ line was used to estimate disk mass of GM Aur. The HD based disk mass is $2.5 - 20.4 \times 10^{-2}\text{M}_\odot$ \citep{mcclure16}, in line with our estimate of 0.118M$_\odot$. Finally, the non-detection of HD in HD163296 \citep{kama20} translates into an upper limit for the disk mass of $0.067\text{M}_\odot$, which is almost half of the value we obtained in this work. 

Another reliable method to trace the disk mass uses the N$_2$H$^+$. This molecule is a chemical tracer of CO-poor gas and can be used to measure the CO-H$_2$ ratio and calibrate CO-based gas masses. By combining N$_2$H$^+$ with C$^{18}$O, \cite{trapman22} estimated disk masses of three protoplanetary disks, including GM Aur. The value they obtained, $1.5 - 9.6 \times 10^{-2}\text{M}_\odot$, is slightly higher compared to our estimate, but it is in an overall good agreement. This method has also been used to probe disk masses of protoplanetary disks in the Lupus star-forming region \citep{anderson22}.

In this context, it is worth mentioning observations of the $^{13}$C$^{17}$O, a very rare CO isotopologue. \cite{booth19} observed this molecule in HD 163296 and this allows for a precise disk mass measurement to be obtained. These authors found that the disk mass that is better at reproducing observations is $M_d = 0.31\text{M}_\odot$, which is discrepant with our inferred value.

As for the dust, its ability to trace the mass is discussed in the next subsection.

\subsection{Gas-to-dust ratio}

With the knowledge of the disk mass, it is possible to evaluate the gas-to-dust ratio, using Eq. \eqref{dustmass} for the dust mass. The results are shown in  Table \ref{gtdratio}. We found values between $100-250$, within  a factor of 2 from the usually assumed valued of 100. This is surprisingly, due to the assumptions we made to obtain the dust mass. Indeed, as we have already mentioned, the optically thin hypothesis for dust emission could lead to a difference of a more than a factor of 2 in the dust mass calculation \citep{guidi16}, underestimating it. In addition, the dust opacity could also vary of a factor of $\sim10$, depending on the grain size and composition. Hence, it is significant overall that the inferred gas-to-dust ratio is so close to the standard value. As for AS 209, we estimated an upper limit for this quantity. Indeed, according to \cite{Veronesiprep}, the minimum measurable mass with the rotation curve is 5\% of the star mass. Taking this value as an upper limit for AS 209 disk mass, it is possible to give an upper limit for the gas-to-dust ratio.

\begin{table}
\caption{Continuum fluxes at 283 GHz, dust masses from Eq. \eqref{dustmass} and gas-to-dust ratio using the best-fit value of the disk mass of the stratified model.}\label{gtdratio}
\resizebox{0.455\textwidth}{!}{
\begin{tabular}{llll}
 & $F_{283}$ [mJy] & $M_\text{dust}$ [M$_\odot$] & Gas-to-dust ratio\\ 
 \hline\\
\textbf{MWC 480} & 943.51 & 0.00138& 108 \\
\textbf{IM Lup} & 536.25 & 0.00075 & 134  \\ 
\textbf{GM Aur} & 347.95 & 0.00049 & 240 \\ 
\textbf{HD 163296} & 1127.97 & 0.00064 & 202  \\ 
\textbf{AS 209} & 414.83 & 0.00034 & < 192 \\ 
\end{tabular}
}
\end{table}
\begin{comment}
\begin{table}[]    
\caption{Flux based radii measurements of $^{12}$CO and $^{13}$CO from \cite{MAPSIV}, ratio between the flux based radii and the best-fit $R_c$ of the stratified model and theoretical value of the flux based radius of the $^{12}$CO from the model of \cite{trapman23}.}
    \centering\resizebox{0.475\textwidth}{!}{
    \begin{tabular}{llllll}
    & $R_{90}^\text{CO}$ & $R_{90}^\text{13CO}$ &$R_{90}^\text{CO}/R_c$ & $R_{90}^\text{13CO}/R_c$ & $R_{90}^\text{Tr}$\\ 
    \hline \\
\textbf{MWC 480}  & $573 \pm 7$ & $149 \pm 7$& 4.48& 3.27& 606 \\
\textbf{IM Lup}  & $753\pm6$ & $540\pm7$ & 6.55 &  4.70& 576 \\
\textbf{GM Aur}  & $616\pm9$& $427\pm9$ &6.42 & 4.45 & 519\\
\textbf{HD 163296}  & $459\pm4$ & $364\pm4$ & 5.03 & 4.00 & 479  \\
\textbf{AS 209}  & $272\pm4$ & $195\pm5$& 2.16 & 1.55 & 226\\
\end{tabular}
}
    \label{radii_table}
\end{table}
\end{comment}
\subsection{Toomre Q}
\begin{figure}
    \centering
    \includegraphics[scale=0.45]{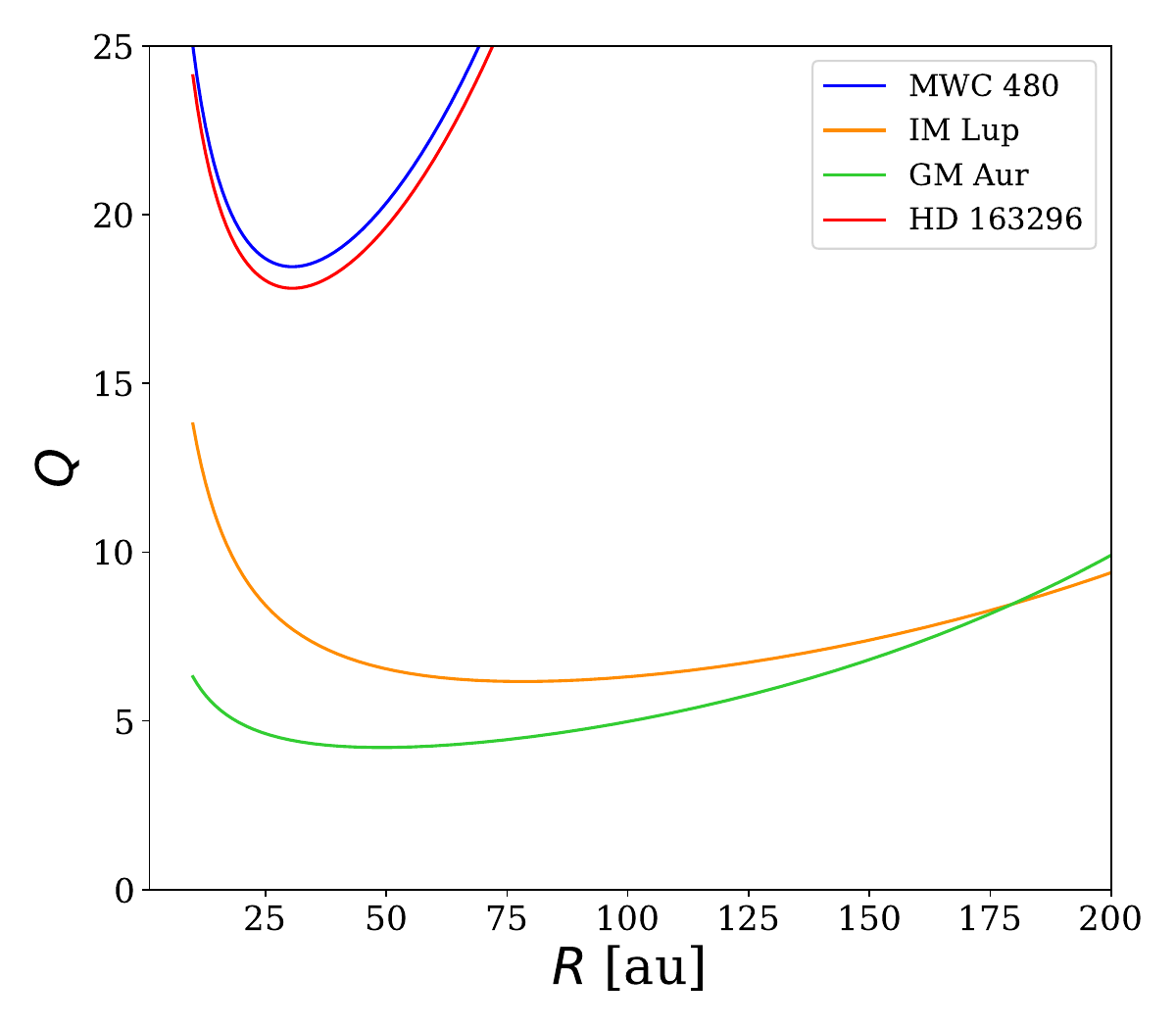}
    \caption{Toomre Q parameter of the MAPS disks with the best-fit parameters of the stratified model. We excluded AS 209 because its best-fit disk mass is compatible with zero.}
    \label{toomreq}
\end{figure}

To investigate the presence of gravitational instability, we used
our best-fit parameters for the stratified model to compute the Toomre parameter \citep{toomre64} which, based on the hypothesis of nearly Keplerian disk ($\kappa \simeq \Omega$), is
\begin{equation}
    Q\simeq\frac{c_\text{s}\Omega}{\pi G\Sigma} = 2\left.\frac{H}{R}\right|_\text{mid}\frac{M_\star}{M_d}\left(\frac{R}{R_c}\right)^{-1}\exp\left[ \frac{R}{R_c}\right],
\end{equation}
where we used Eq. \eqref{eq: surf density} for the surface density. According to the WKB quadratic dispersion relation \citep{lin64,toomre64}, the onset of the instability happens when $Q\sim 1$. Figure \ref{toomreq} shows the profile of the Q parameter for the MAPS sample, except for AS209, since its disk mass estimate is compatible with zero. Every disk is gravitationally stable, according to the Toomre criterion, since $Q>1$. Interestingly, the two disks that showed spiral structures (IM Lup and GM Aur) have a Toomre profile that is lower than the others, with a minimum value of $\sim4$ for GM Aur and $\sim 6$ for IM Lup. \cite{lau78} showed that a WKB description of gravitational instability can still be obtained under less restrictive conditions compared to the quadratic relation. They showed that disks that are locally stable according to the Q criterion might still generate large scale spiral waves. In general, other mechanisms could increase the critical value of the Toomre parameter, such as external irradiation \citep{lin16,lohnert20} or dust-driven gravitational instability \citep{longarini23a,longarini23b}. Hence, we do not exclude the possibility that gravitational instability is at play in GM Aur and IM Lup.

\section{Conclusions}\label{concl}
The kinematic data of protoplanetary disks display velocity differences between $^{12}$CO and $^{13}$CO that cannot be explained through a vertically isothermal model, given the systematic shift between rotation curves of CO isotopologues. In this work, we predict how thermal stratification affects the density and the velocity field of a protoplanetary disk. 

We used SPH simulations to test our model, finding excellent agreement, and then we applied it to the MAPS sample. We extracted the rotation curves of CO isotopologues ($^{12}$CO and $^{13}$CO) and we carried out fitting for the star mass, disk mass, and scale radius - both with a vertically isothermal and a stratified model. The quality of the fit significantly improves when thermal stratification is taken into account and the best-fit parameters are more realistic and aligned with literature. All our results are summarized in  Table \ref{fits}.

Typically, when thermal stratification is considered, the best-fit value for the star mass tends to rise. This can be intuitively understood, as an isothermal model would favor a star mass that lies between that of $^{13}$CO and $^{12}$CO; its  underestimation is due to the slower rotation of $^{12}$CO. Conversely, the stratified model encapsulates the difference between the two curves, mitigating the underestimation issue and resulting in a more accurate mass estimate. While an isothermal model provides a satisfactory fit at small radii, the fit worsens at large radii where the difference between $^{12}$CO and $^{13}$CO is larger. The fit tries to compensate for this by increasing the disk mass, most of which resides at large radii, thereby changing the predicted curve only in the outer parts of the disk. Ultimately, a more accurate description of the thermal structure through a stratified model leads to a realistic estimate of the scale radius.

%We note that the inclusion of thermal stratification into our model enhances our comprehension of the observed data within protoplanetary disks. This addition leads to more accurate estimate of disk properties, resulting in improved $\chi^2$ values across all systems under examination.
We note that the inclusion of the vertical gradient of temperature into our model results in improved $\chi^2$ values across all systems under examination.
This work demonstrates the impact of thermal 
stratification in the disk dynamics and highlights the importance of 
having a precise knowledge of the disk temperature structure to infer 
physical quantities such as the stellar mass, the disk mass and the disk scale radius in a meaningful manner. Thanks to the generality of our calculations, we can study the density and velocity profile of thermally stratified disk according different prescription of temperature simply.

\begin{acknowledgements}
This paper makes use of the following ALMA data: ADS/JAO.ALMA\#2018.1.01055.L. ALMA is a partnership of ESO (representing its member states), NSF (USA), and NINS (Japan), together with NRC (Canada), NSC and ASIAA (Taiwan), and KASI (Republic of Korea), in cooperation with the Republic of Chile. The Joint ALMA Observatory is operated by ESO, AUI/NRAO, and NAOJ. This work has received funding from the European Union’s Horizon 2020 research and innovation programme under the Marie Sklodowska-Curie grant agreement \# 823823 (RISE DUSTBUSTERS project). G.R. is funded by the European Union under the European Union’s Horizon Europe Research \& Innovation Programme No.~101039651 (discEvol) and by the Fondazione Cariplo, grant no. 2022-1217. S.F. is funded by the European Union (ERC, UNVEIL, 101076613). Views and opinions expressed are however those of the author(s) only and do not necessarily reflect those of the European Union or the European Research Council. Neither the European Union nor the granting authority can be held responsible for them. S.F. acknowledges financial contribution from PRIN-MUR 2022YP5ACE. CH is funded by a Research Training Program Scholarship from the Australian Government, and acknowledges funding from the Australian Research Council via DP220103767. MB and JS have received funding from the European Research Council (ERC) under the European Union’s Horizon 2020 research and innovation programme (PROTOPLANETS, grant agreement No. 101002188).
The authors thank Pietro Curone and Claudia Toci for useful discussions. 
\end{acknowledgements}

\bibliographystyle{aa} % style aa.bst
\bibliography{bibliography.bib} % your references Yourfile.bib
\nocite{*}

\begin{appendix}
\section{Computing the pressure gradient}
\label{calc}
Assuming a barotropic fluid, the pressure contribution to the rotation curve is
\begin{equation}
    \frac{R}{\rho}\frac{dP}{dR}\Bigg|_z= \frac{R}{\rho}\Bigg(c_\text{s}^2\frac{d\rho}{dR}\Bigg|_z+\rho\frac{dc_\text{s}^2}{dR}\Bigg|_z\Bigg),
\end{equation}
which can also be expressed as:
\begin{equation}
\begin{split}
\label{appendix: grad press}
    \frac{R}{\rho}\frac{dP}{dR}\Bigg|_z = c_\text{s}^2\frac{R}{\rho_\text{mid}g}&\Bigg(\rho_\text{mid}\frac{dg}{dR}\Bigg|_z+g\frac{d\rho_\text{mid}}{dR}\Bigg|_z\Bigg)+\\
    R&\Bigg(c_{\text{s,mid}}^2\frac{df}{dR}\Bigg|_z+f\frac{dc_{\text{s,mid}}^2}{dR}\Bigg|_z\Bigg) =\\
    c_{\text{s}}^2\Bigg[-\gamma'&- (2-\gamma)\left(\frac{R}{R_c}\right)^{2-\gamma}+R\frac{d\log(fg)}{dR}\Bigg|_z\Bigg].
\end{split}
\end{equation}
Assuming hydrostatic equilibrium in the vertical direction, we can write an explicit expression for the last term in Eq.\eqref{appendix: grad press}:
\begin{equation}
\begin{split}
     -\Omega_\text{k}^2 z\left[1+\left(\frac{z}{R}\right)^2\right]^{-3/2} &= -\frac{d\Phi_\star}{dz} = \frac{1}{\rho}\frac{dP}{dz} = \\
     \frac{1}{\rho}P_\text{mid}\frac{d(fg)}{dz} &= c_{\text{s,mid}}^2f\frac{d\log(fg)}{dz}
\end{split}
\end{equation}
and, assuming $f(z=0)=1$ (therefore $T(R,z=0)=T_\text{mid}(R))$,
\begin{equation}  
    \log(fg) = -\frac{1}{H_\text{mid}^2}\int_0^z \frac{z'}{f}\left[1+\left(\frac{z'}{R}\right)^2\right]^{-3/2} dz'.
\end{equation}
Deriving with respect to R we obtain:
\begin{equation}
\begin{split}
\label{derivlog}
    &R\frac{d\log(fg)}{dR}\Bigg|_z= R\Bigg[\frac{2}{H_\text{mid}^3}\frac{dH_\text{mid}}{dR}\int_0^z \frac{1}{f}\frac{z'dz'}{\Big(1+(z'/R)^2\Big)^{3/2}}+\\
    &\frac{1}{H_\text{mid}^2}\int_0^z \frac{1}{f^2}\frac{df}{dR}\frac{z'dz'}{\Big(1+(z'/R)^2\Big)^{3/2}}
    -\frac{1}{H_\text{mid}^2}\int_0^z \frac{1}{f}\frac{3z'^3/R^3dz'}{\Big(1+(z'/R)^2\Big)^{5/2}}\Bigg] =\\
     &-\frac{q}{H_\text{mid}^2}\int_0^z \frac{1}{f}\frac{z'dz'}{(1+z'^2/R^2)^{3/2}} +
    \frac{R}{H_\text{mid}^2}\int_0^z \frac{\dot{f}}{f^2}\frac{z'dz'}{(1+z'^2/R^2)^{3/2}}+\\
    &\frac{3}{H_\text{mid}^2}\int_0^z\frac{1}{f}\frac{z'dz'}{(1+z'^2/R^2)^{5/2}},
\end{split}
\end{equation}
where $\cdot=d/dR$. Expanding the logarithmic term is useful since it is now clear that the azimuthal velocity depends both on the temperature $(f)$ and on the temperature gradient $(\dot{f})$ along the radial direction. Moreover, it is easy to see the contribution of the vertical thermal stratification: in our model $f$ and $\dot{f}$ are functions of $z$ and thus they have to be integrated to compute azimuthal velocity; whereas in vertically isothermal models, this does not happen, since $f=1$ and $\dot{f}=0$. 
We note that it is not analytically possible to determine generally the sign of Eq. \eqref{derivlog} and, thus, the effect that we see on the pressure gradient.
However, for all the considered cases it never overcomes the gravitational contribution and it does lead to a faster deceleration in the rotation.

\section{Dullemond prescription}
\label{Dullcorrection}
According to \citet{MAPSIV}, the temperature prescription given by \citet{Dullemond} (Eq.\eqref{Dull}) fits  the data well, but we note that in this case, $f(R, z=0)\neq 1$ and  the temperature does not smoothly connect to its value at midplane since $T(R,z=0)\neq T_\text{mid}(R)$. 
To evaluate this discrepancy, considering $ T_\epsilon(R)\approx T_\text{mid}(R)$, we compute:
\begin{equation}\label{estimate}
\begin{split}
    \Delta(R)&= 1- \frac{T(R,0)}{T_\text{mid}} = 1-f(R,z=0)\approx\\
    &-\frac{1}{8}\Bigg(\frac{T_\text{atm}}{T_\text{mid}}\Bigg)^{4}(R)(1-\tanh\alpha) =\\
    &- \frac{1}{8}\Bigg(\frac{T_{\text{atm},100}}{T_{\text{mid},100}}\Bigg)^4 \Bigg(\frac{R}{100\text{au}}\Bigg)^{-4(q_\text{atm}-q)}(1-\tanh\alpha),
\end{split}
\end{equation}
which is strongly dependent on $T_\text{mid}, T_\text{atm}$ and $\alpha$. As shown in Table \ref{maps_params}, the deviation of the actual midplane temperature from $T(R,z=0)$ as computed from the prescription given by \citet{Dullemond} is $\lesssim10\%$ in our regions of interest ($R>100$au). This discrepancy could have relevance for systems such as GM Aur, HD 163296 and MWC 480. To examine this further, we conducted fits for these systems using both $T_\text{mid}(R)$ and $T(R,z=0)$ as the midplane temperature. It was observed that this choice does not significantly alter the results.
Thus, $T_\text{mid}\approx T(T,z=0)$ and $f(R,z=0)\approx1$. We can consider Eqs. \eqref{Dull} and \eqref{Dull f} as a good parameterization for the temperature.

%As shown in Table \ref{maps_params}, for the five disks studied by the  MAPS large program the discrepancy of the actual midplane temperature from the value predicted by \citet{Dullemond} prescription at $z=0$au is $\lesssim10\%$ in the outer part of the disk ($R>100$au) and thus we can approximate $T_\epsilon\simeq T_\text{mid}$ and consider equations \eqref{Dull} and \eqref{Dull f} as a good parametrization for the temperature. 

\begin{table}\caption{Values of parameters used in the estimate \eqref{estimate}.}\label{maps_params}
\resizebox{0.5\textwidth}{!}{
\begin{tabular}{lllll}
 &{$(T_\text{atm}/T_\text{mid})_{100}$} & {$q_\text{atm}-q$} &{$\alpha$}  &{$|\Delta(R=100$au)|}\\ \hline \\
\textit{IM Lup} &$1.44$ &$0.05$ &$4.91$  &$5.8 \cdot10^{-5}$ \\
\textit{GM Aur} &$2.4$ &$0.54$ &$2.57$  &$4.8 \cdot10^{-2}$\\
\textit{AS 209} &$1.48$ &$0.41$ &$3.31$  &$1.6 \cdot10^{-3}$ \\
\textit{HD 163296} &$2.63$ &$0.43$ &$3.01$ &$2.9 \cdot10^{-2}$\\
\textit{MWC 480} &$2.56$ &$0.47$ &$2.78$  &$4.1 \cdot10^{-2}$\\
\end{tabular}
}
\end{table}

%\begin{tabular}{llllll}
% &{$(T_\text{atm}/T_\text{mid})_{100}$} & {$q_\text{atm}-q$} &{$\alpha$}  &{$|\Delta(R=100$au)|} &{$|\Delta(R=200$au)|}\\ \hline \\
%\textit{IM Lup} &$1.44$ &$0.05$ &$4.91$  &$5.8 \cdot10^{-5}$ &$5\cdot10^{-5}$\\
%\textit{GM Aur} &$2.4$ &$0.54$ &$2.57$  &$4.8 \cdot10^{-2}$ &$1\cdot10^{-2}$\\
%\textit{AS 209} &$1.48$ &$0.41$ &$3.31$  &$1.6 \cdot10^{-3}$ &$5\cdot10^{-4}$\\
%\textit{HD 163296} &$2.63$ &$0.43$ &$3.01$ &$2.9 \cdot10^{-2}$ &$9\cdot10^{-3}$\\
%\textit{MWC 480} &$2.56$ &$0.47$ &$2.78$  &$4.1 \cdot10^{-2}$ &$1\cdot10^{-2}$\\

\section{Corner plots}\label{app_corner}
\begin{figure*}
    \centering
    \includegraphics[scale=0.385]{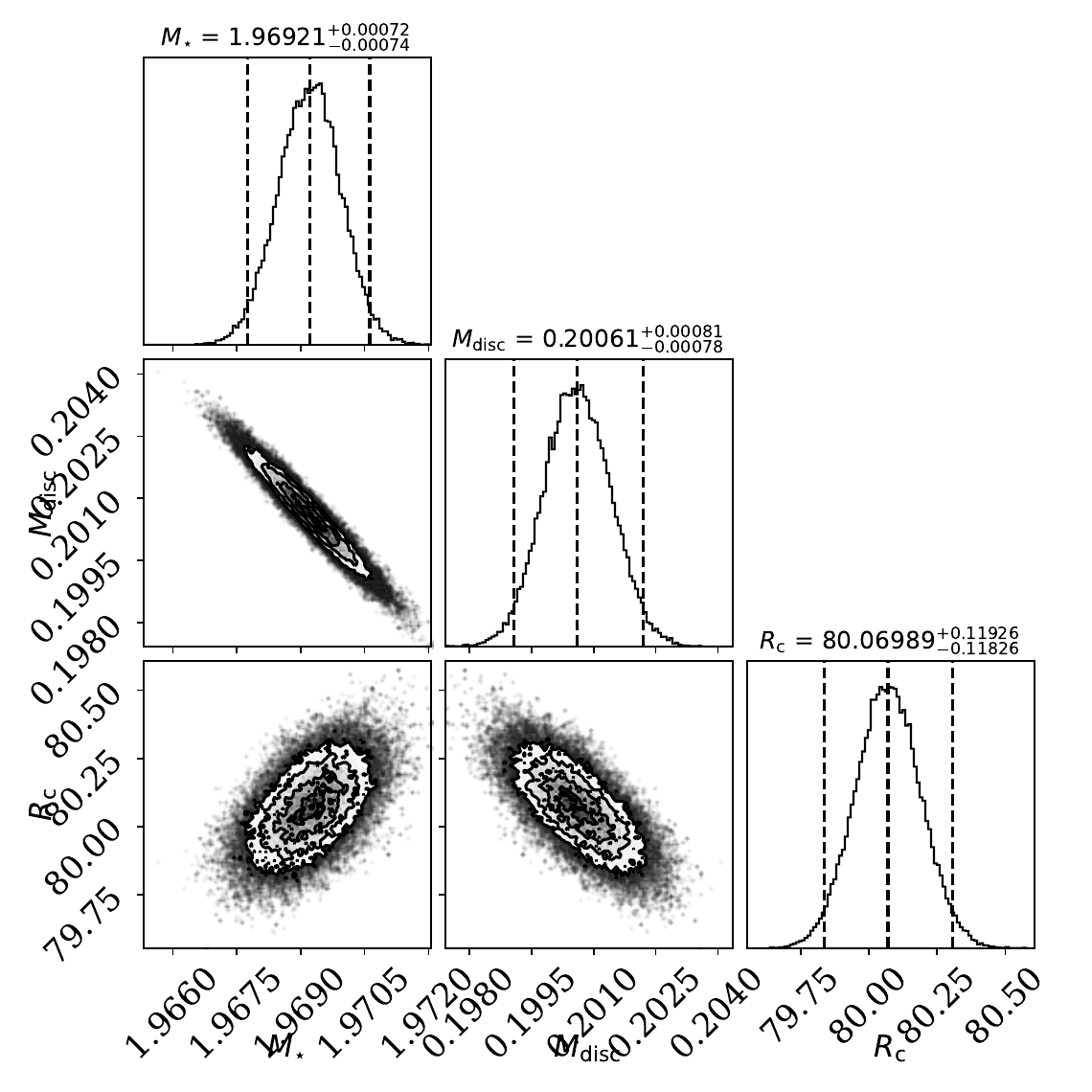}
    \includegraphics[scale=0.385]{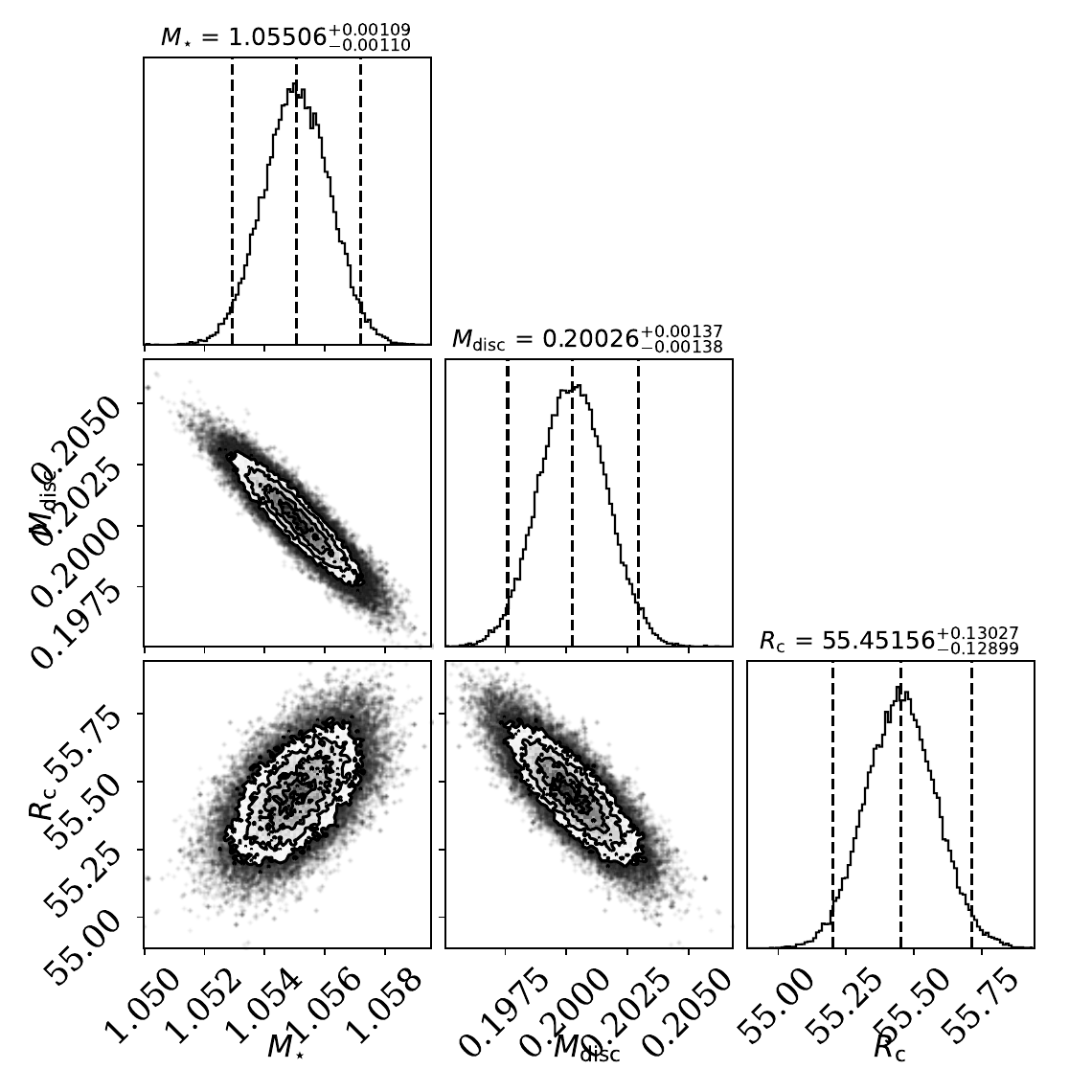}
    \includegraphics[scale=0.385]{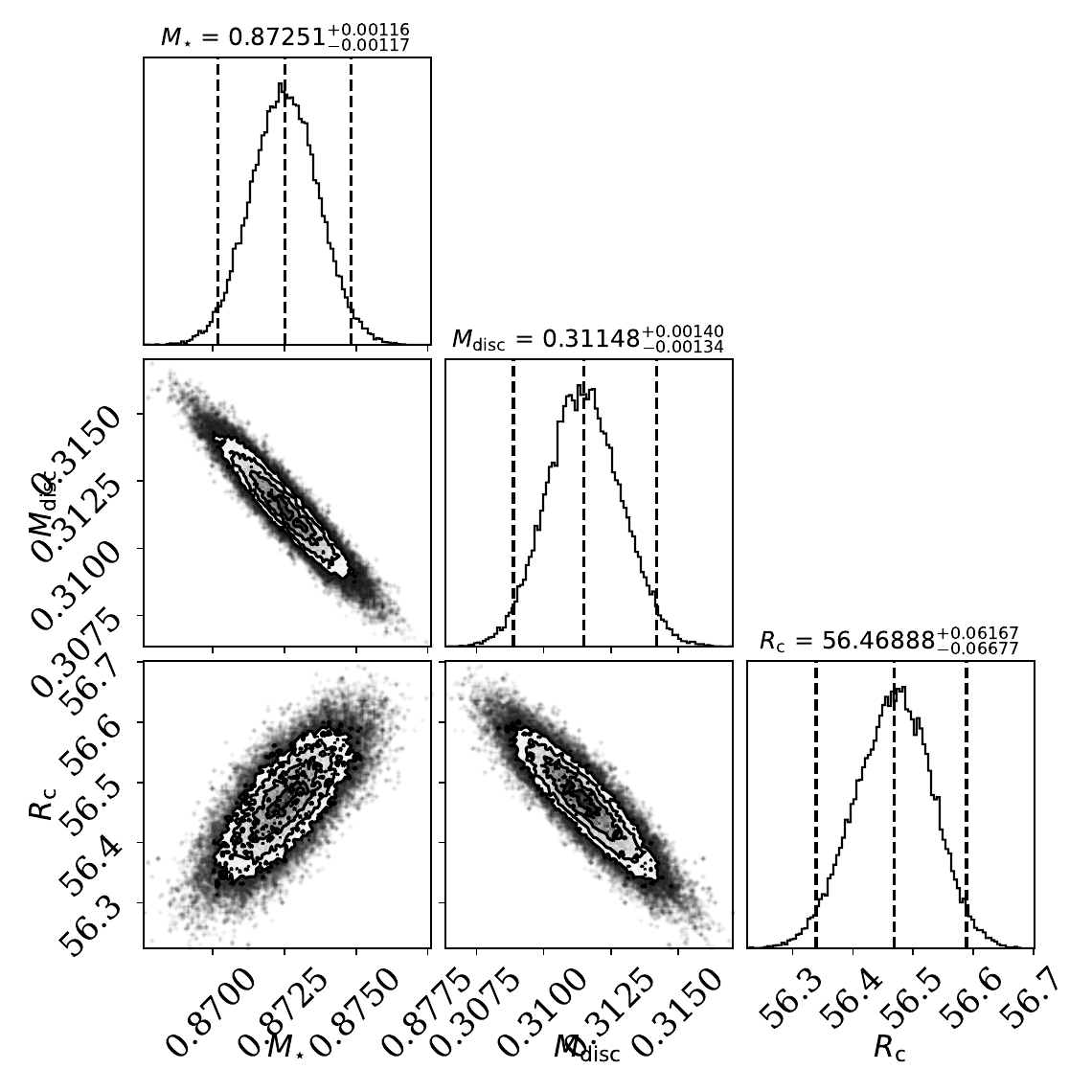}
    \includegraphics[scale=0.385]{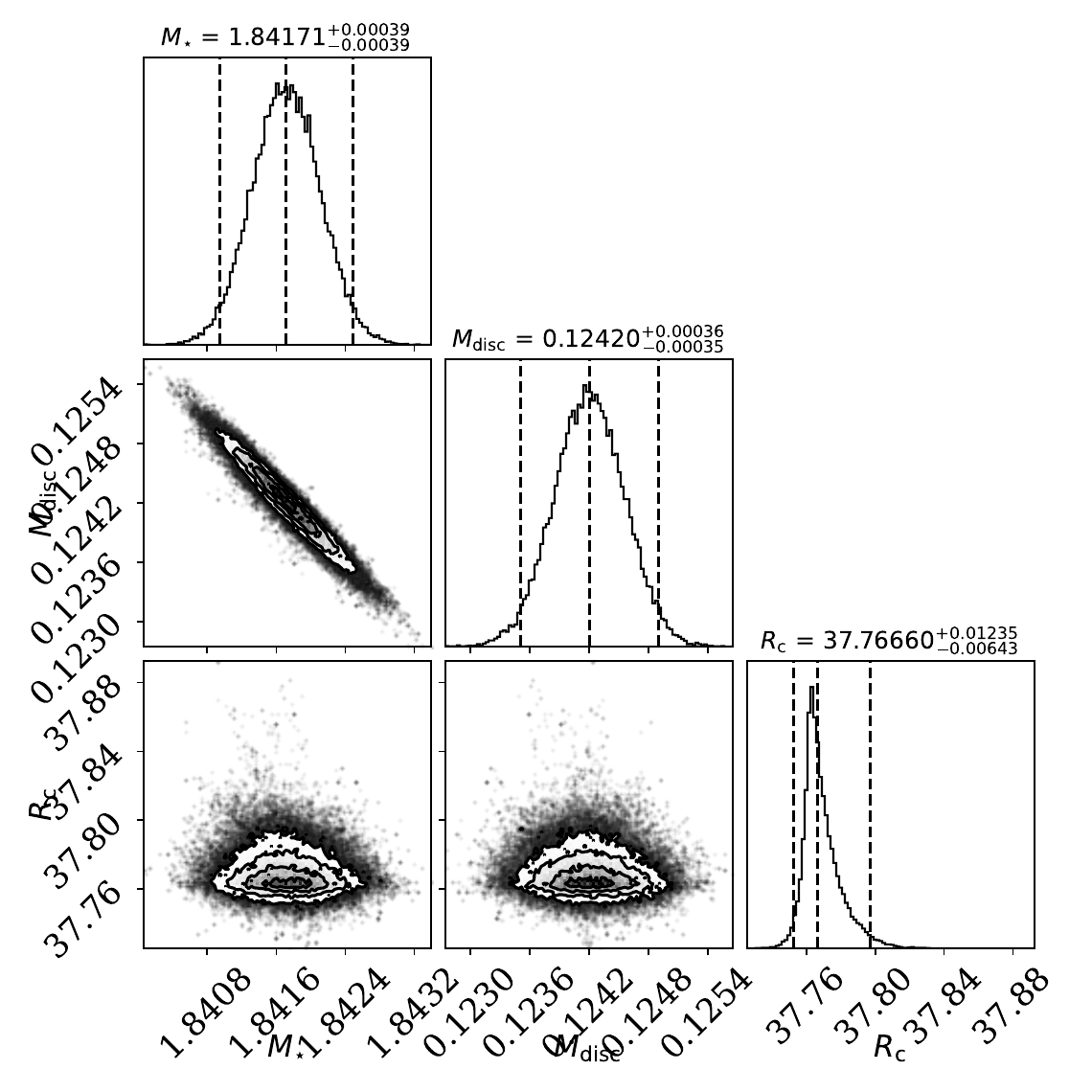}
    \includegraphics[scale=0.385]{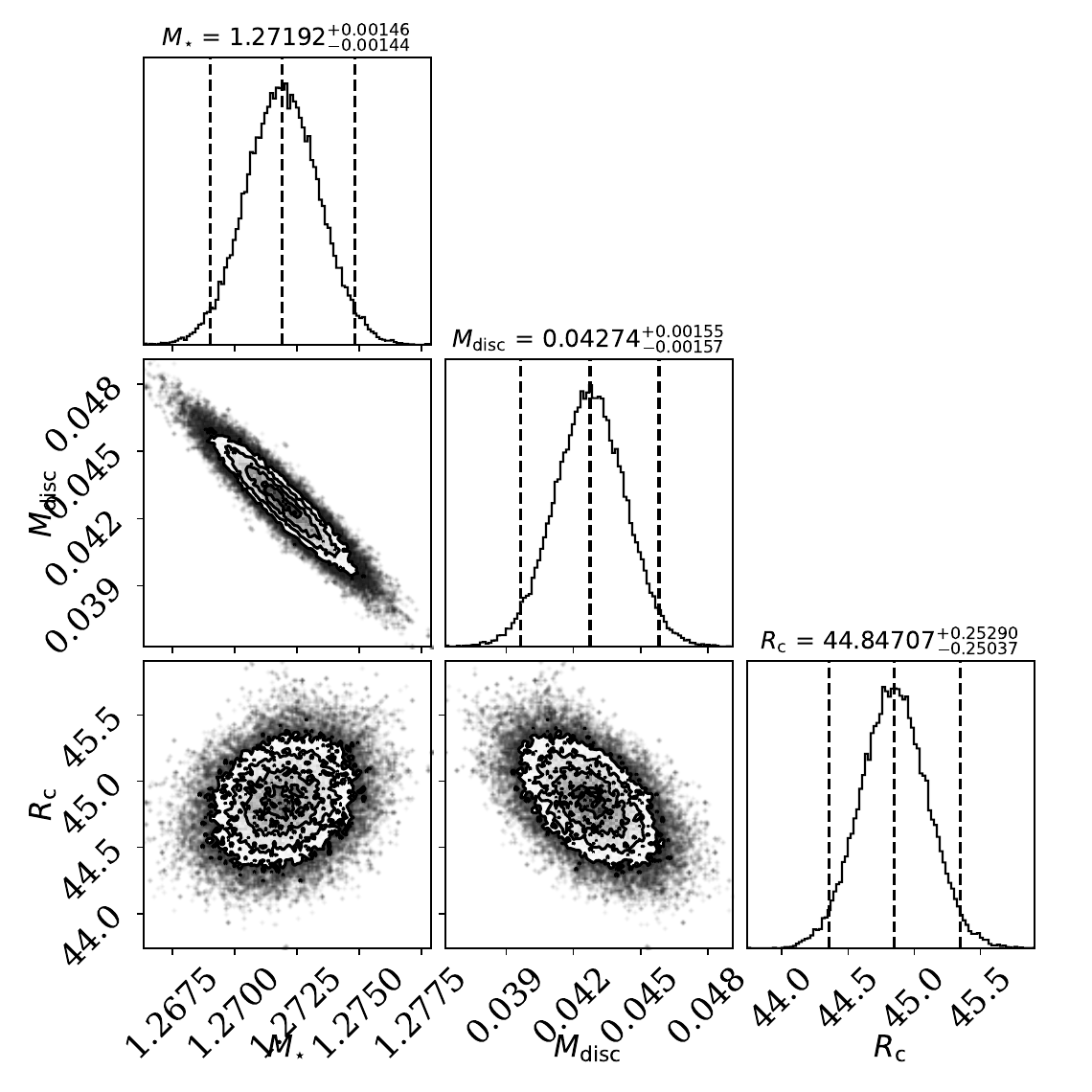}
    \caption{Corner plots of the MCMC fitting procedure according the vertically isothermal model. They show the distribution of the three relevant fitting parameters for the five disks of the MAPS large program. From top left to bottom: MWC 480, IM Lup, GM Aur, HD 163296, and AS
209. }
    \label{corner_iso}
\end{figure*}

\begin{figure*}
    \centering
    \includegraphics[scale=0.385]{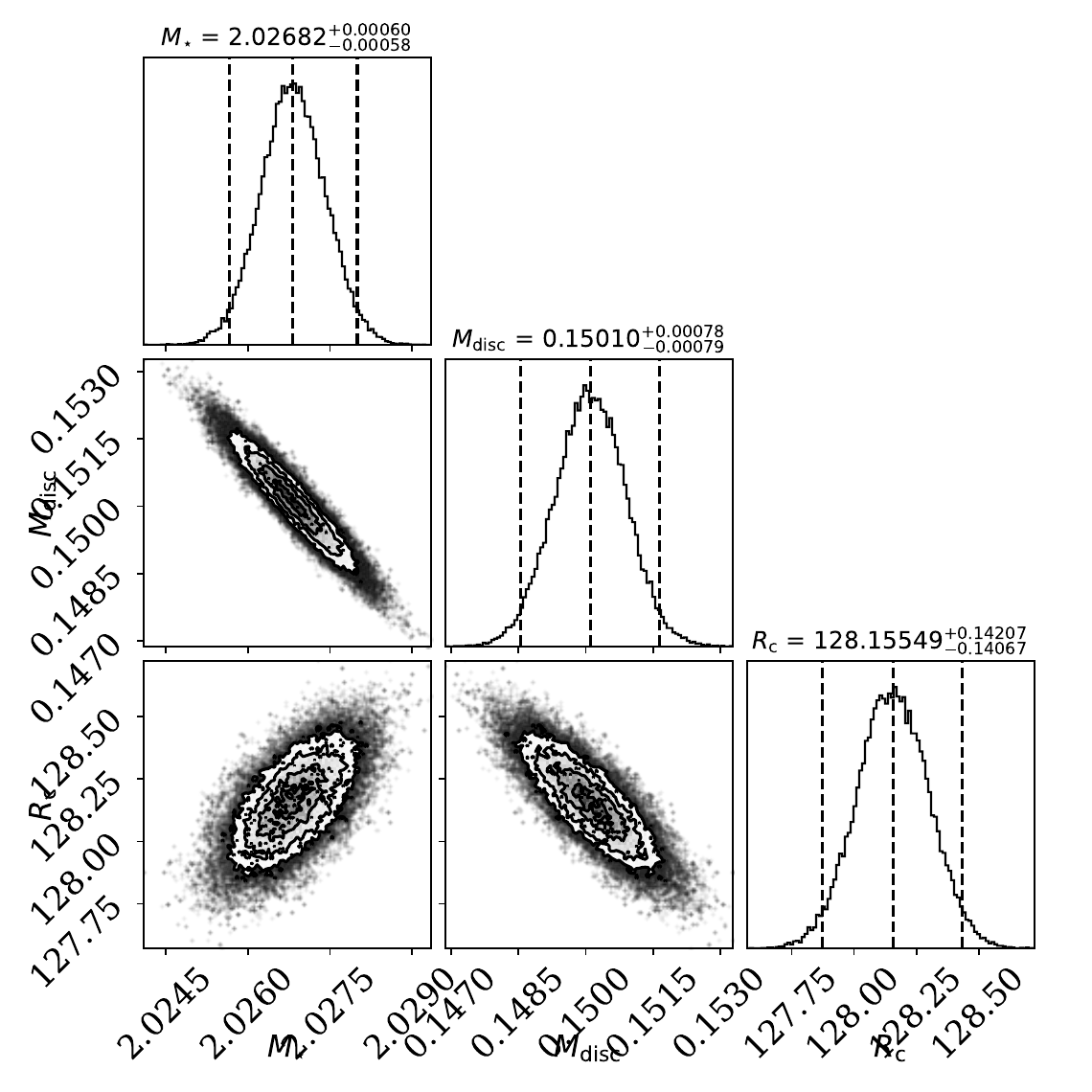}
    \includegraphics[scale=0.385]{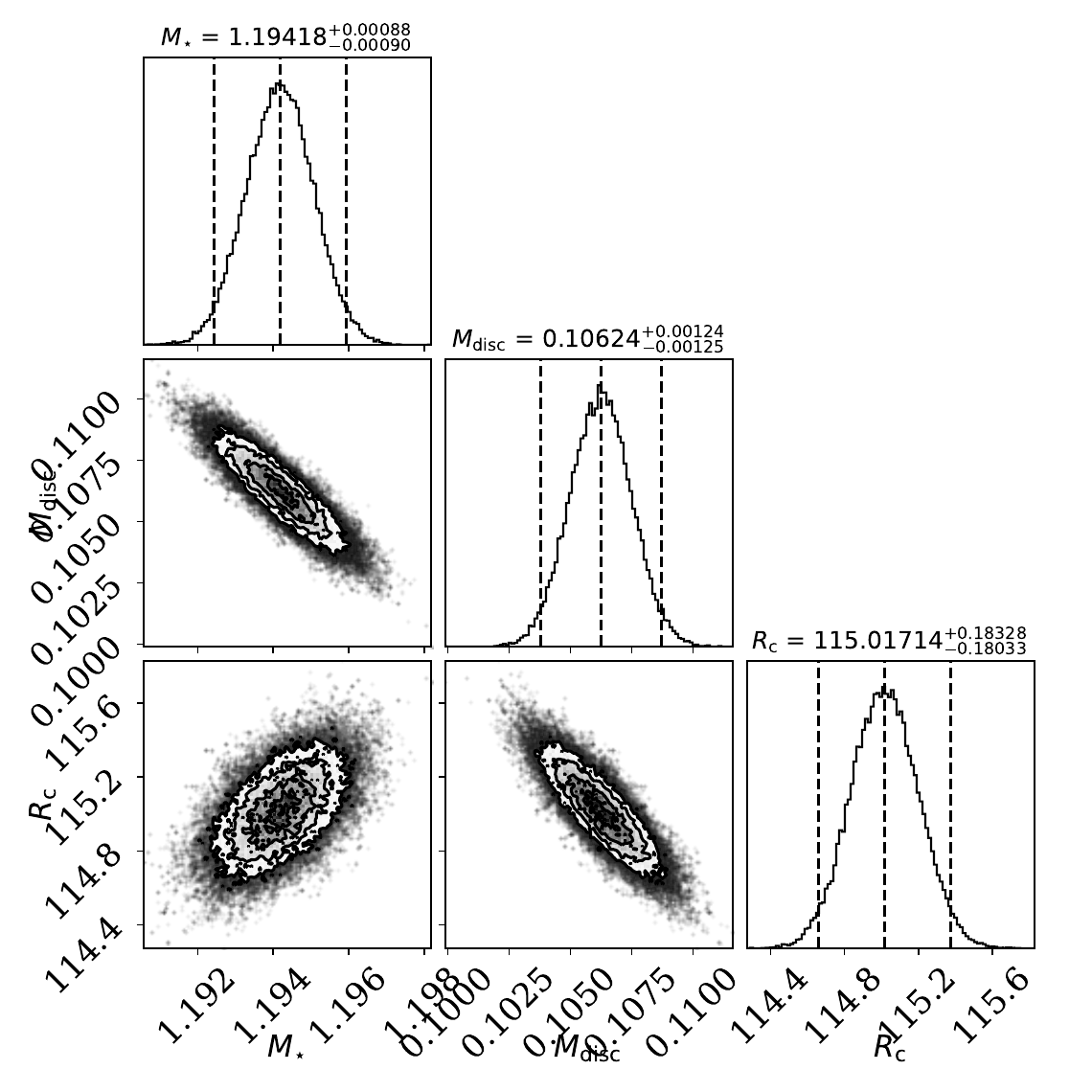}
    \includegraphics[scale=0.385]{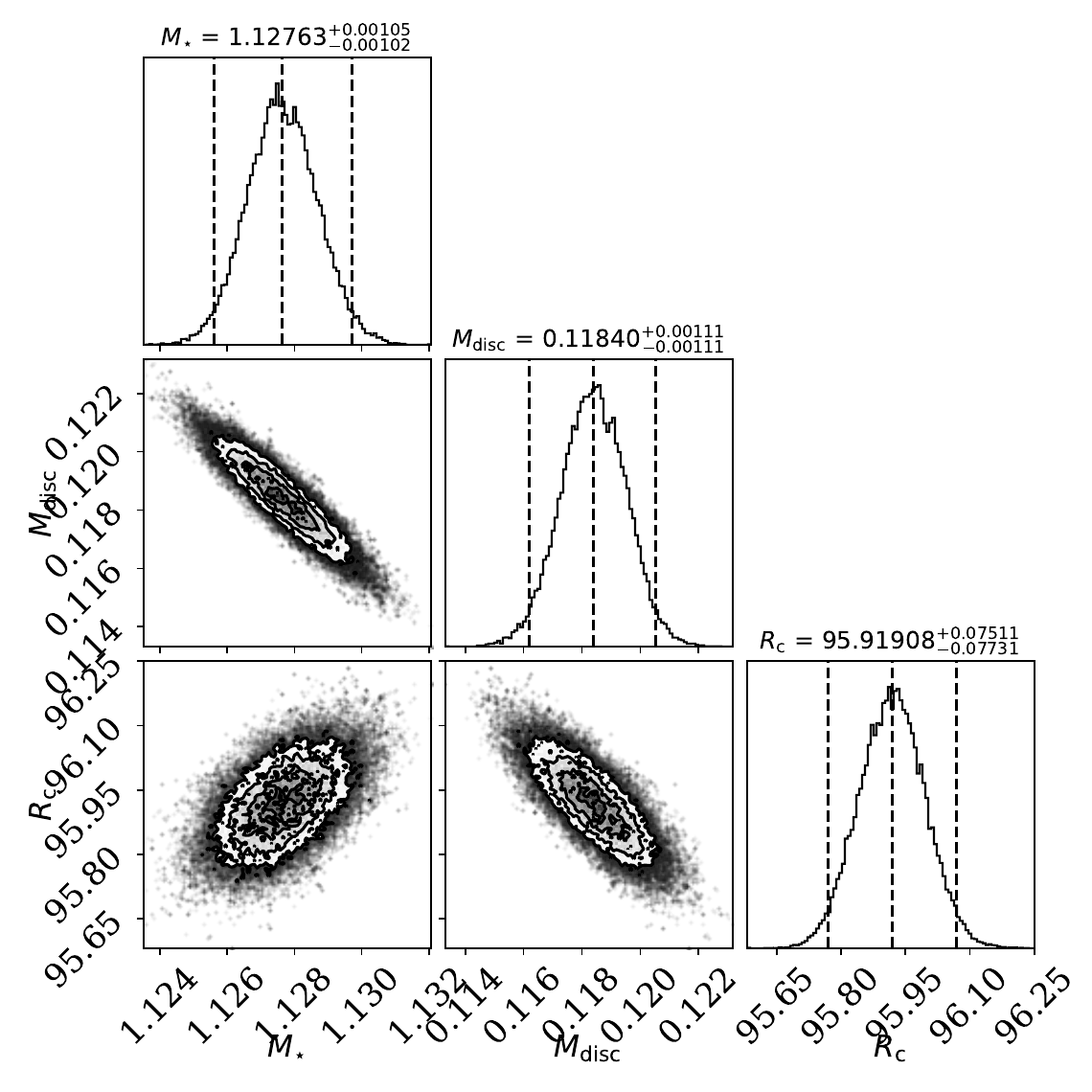}
    \includegraphics[scale=0.385]{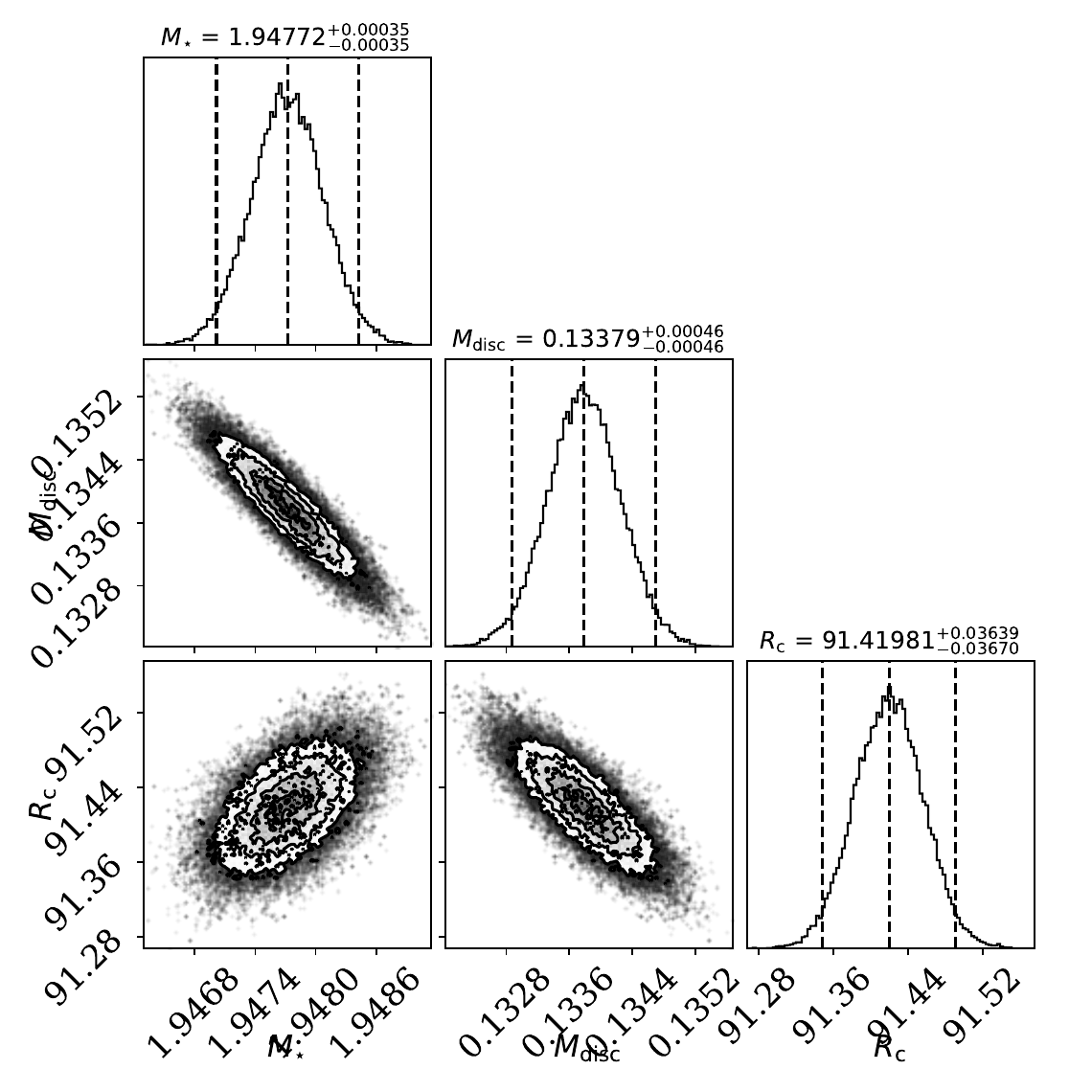}
    \includegraphics[scale=0.385]{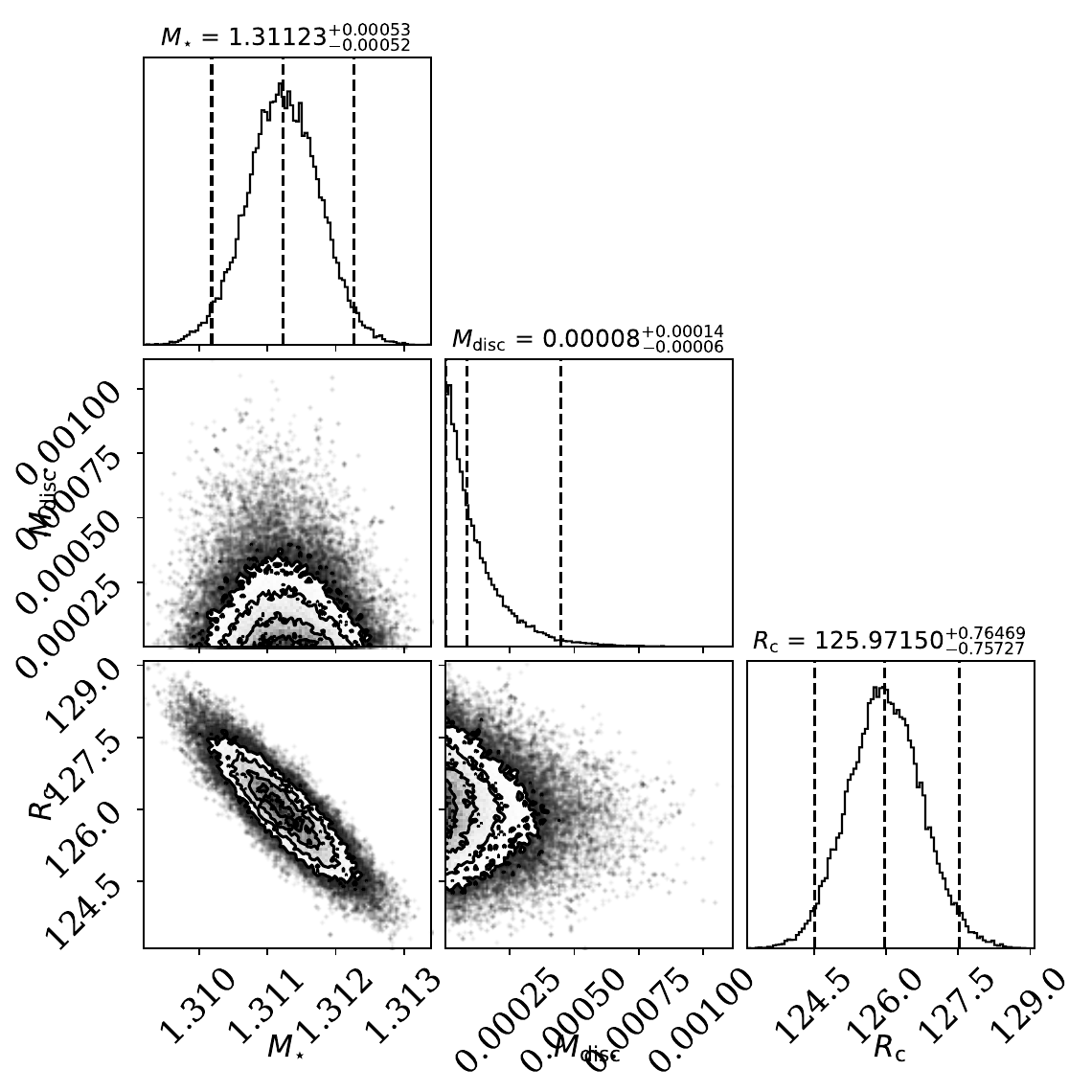}
    \caption{Corner plots of the MCMC fitting procedure according the stratified model. They show the distribution of the three relevant fitting parameters for the five disks of the MAPS large program. From top left to bottom: MWC 480, IM Lup, GM Aur, HD 163296, and AS
209.}
    \label{corner_strat}
\end{figure*}

In Figs \ref{corner_iso} and \ref{corner_strat}, we present the corner plots of the MCMC fitting procedure for the studied parameters under, respectively, the vertically isothermal and the stratified model.

\end{appendix}

\end{document}